\documentclass[12pt]{article}
\usepackage{amsmath}
\usepackage{amssymb}
\usepackage{epsfig}
\usepackage{pslatex}

\makeatletter

\@addtoreset{equation}{section}
\makeatother
\newcommand{\bea}{\begin{eqnarray}}
\newcommand{\eea}{\end{eqnarray}}
\newcommand{\be}{\begin{eqnarray}}
\newcommand{\ee}{\end{eqnarray}}

\def\det{\mathop{\rm det}}

\def\rmd{{\rm d}}

\newcommand{\del}{\partial}
\DeclareMathOperator*{\Tr}{{\rm Tr}}
\DeclareMathOperator*{\tr}{{\rm tr}}

\newcommand{\Yb}{Y^{\dag}}
\newcommand{\psib}{\psi^{\dag}}

\begin{document}

\begin{titlepage}
\vskip-0cm
\begin{flushright}
SNUST 080903\\
{\tt arXiv:0809.3786[hep-th]}
\end{flushright}
\vskip0.5cm
\centerline{\Large \bf  Wilson Loops in Superconformal Chern-Simons Theory}
\vskip0.25cm
\centerline{\Large \bf and}
 \vskip0.25cm
 \centerline{\Large \bf Fundamental Strings in Anti-de Sitter Supergravity Dual}
\vspace{1.25cm}
\centerline{\large Soo-Jong Rey, \,\,\, Takao Suyama, \,\,\, Satoshi Yamaguchi}
\vspace{1.25cm}
\centerline{\sl School of Physics \& Astronomy, Seoul National University, Seoul 151-747 {\rm KOREA}}
\vskip0.25cm
\centerline{\tt sjrey@snu.ac.kr \,\,\,\,\, suyama, \, yamaguch@phya.snu.ac.kr}
\vspace{1cm}
\centerline{ABSTRACT}
\vspace{1.0cm}
\noindent
We study Wilson loop operators in three-dimensional, ${\cal N}=6$ superconformal Chern-Simons theory dual to IIA superstring theory on AdS$_4 \times \mathbb{CP}^3$. Novelty of Wilson loop operators in this theory is that, for a given contour, there are two linear combinations of Wilson loop transforming oppositely under time-reversal transformation. We show that one combination is holographically dual to IIA fundamental string, while orthogonal combination is set to zero. We gather supporting evidences from detailed comparative study of generalized time-reversal transformations in both D2-brane worldvolume and ABJM theories. We then classify supersymmetric Wilson loops and find at most ${1 \over 6}$ supersymmetry. We next study Wilson loop expectation value in planar perturbation theory. For circular Wilson loop, we find features remarkably parallel to circular Wilson loop in ${\cal N}=4$ super Yang-Mills theory in four dimensions. First, all odd loop diagrams vanish identically and even loops contribute nontrivial contributions. Second, quantum corrected gauge and scalar propagators take the same form as those of ${\cal N}=4$ super Yang-Mills theory. Combining these results, we propose that expectation value of circular Wilson loop is given by Wilson loop expectation value in pure Chern-Simons theory times zero-dimensional Gaussian matrix model whose variance is specified by an interpolating function of `t Hooft coupling. We suggest the function interpolates smoothly between weak and strong coupling regime, offering new test ground of the AdS/CFT correspondence.

\end{titlepage}
\section{Introduction}
The proposal of holographic principle put forward by Maldacena \cite{Maldacena:1997re} has changed fundamentally the way we understand quantum field theory and quantum gravity. In particular, the AdS-CFT correspondence between ${\cal N}=4$ super Yang-Mills theory and Type IIB superstring on AdS$_5 \times {\mathbb S}^5$, followed by diverse variant setups thereafter, enormously enriched our understanding of nonperturbative aspects of gauge and string theories. In exploring
holographic correspondence between gauge and string theory sides, an important class of physical observable is provided by semiclassical fundamental strings and D-branes in string theory side and by topological defects in gauge theory side. In particular, the Wilson loop operator~\cite{Wilson:1974sk} extended to ${\cal N}=4$ super Yang-Mills theory was proposed and identified with macroscopic fundamental string on AdS$_5 \times \mathbb{S}^5$ \cite{Rey:1998ik, Maldacena:1998im}. During the ensuing development of holographic correspondence between gauge and string theories, the proposal of \cite{Rey:1998ik, Maldacena:1998im} became an essential toolkit for extracting physics from diverse variants of gauge-gravity
correspondence.  Among those further developments, one important step was the observation that the exact expectation value of the ${1 \over 2}$-supersymmetric circular Wilson loop is computable by a Gaussian matrix model \cite{Erickson:2000af,Drukker:2000rr,Pestun:2007rz}.

Recently, Aharony, Bergman, Jafferis and Maldacena (ABJM) \cite{Aharony:2008ug} put forward a new account of the AdS-CFT correspondence: three-dimensional ${\cal N}=6$ superconformal Chern-Simons theory dual to Type IIA string theory on AdS$_4 \times \mathbb{CP}^3$. Both sides of the correspondence are characterized by two integer-valued coupling parameters $N$ and $k$. On the superconformal Chern-Simons theory side, they are the rank of product gauge group U($N) \times \overline{{\rm U}(N)}$ and Chern-Simons levels $+k, -k$, respectively. On the Type IIA string theory side, they are related to spacetime curvature and Ramond-Ramond fluxes, all measured in string unit. Much the same way as the counterpart between ${\cal N}=4$ super Yang-Mills theory and Type IIB string theory on AdS$_5 \times \mathbb{S}^5$, we can put the new correspondence into precision tests in the planar limit:
\bea
N \rightarrow \infty,  \qquad k \rightarrow \infty \qquad
\mbox{with} \qquad \lambda \equiv {N \over k} \quad \mbox{fixed}
   \label{tHooft}
\eea
by interpolating `t Hooft coupling parameter $\lambda$ between superconformal Chern-Simons theory regime at $\lambda \ll 1$ and semiclassical AdS$_4 \times \mathbb{CP}^3$ string theory regime at $\lambda \gg 1$.

The purpose of this paper is to identify Wilson loop operators in the ABJM theory which corresponds to a macroscopic Type IIA fundamental string on AdS$_4 \times \mathbb{CP}^3$ and put them to a test
by studying their quantum-mechanical properties. The proposed Wilson loop operators involve both gauge potential and a pair of bi-fundamental scalar fields, a feature already noted in four-dimensional ${\cal N}=4$ super Yang-Mills theory. Typically, functional form of the Wilson loop operator is constrained severely by the requirement of affine symmetry along the contour $C$, by superconformal symmetry on $\mathbb{R}^{1,2}$, and by gauge and SU(4) symmetries. We shall find that, in the ABJM theory, there are two elementary Wilson loop operators determined by these symmetry requirement:
\bea
W_{\bf N} \, [C, M] &=& {1 \over N} \mbox{Tr} {\cal P} \exp i \oint_C \rmd \tau \Big(A_m \dot{x}^m(\tau) + {M_I}^J(\tau) Y^IY^\dagger_J\Big) \nonumber \\
\overline{W}_{\bf N} [C, M] &=& {1 \over N} \overline{\mbox{Tr}} {\cal P} \exp i \oint_C \rmd \tau \Big(\overline{A}_m \dot{x}^m (\tau) + {M_J}^I(\tau)Y^\dagger_I Y^J\Big).
\eea
We first determine conditions on $x^m(\tau), {M_I}^J(\tau)$ in order for the Wilson loop to keep unbroken supersymmetry. We shall find that there is a unique Wilson loop preserving ${1 \over 6}$ of ${\cal N}=6$
superconformal symmetry. We shall then study vacuum expectation value of these Wilson loops both in planar perturbation theory of the ABJM theory and in minimal surface of the string worldsheet in AdS$_4 \times \mathbb{CP}^3$. We also study determine functional form of ${M_I}^J$ from various symmetry considerations. We shall then propose that the linear combination of Wilson loops:
\bea
{\cal W}_{\bf N} [C, M] := {1 \over 2} \Big( W_{\bf N} [C, M] + \overline{W}_{{\bf N}} [C, M] \Big)
\eea
is identifiable with appropriate Type IIA fundamental string configuration and that the opposite linear combination is mapped to zero.
We gather evidences for these proposal from detailed study for relation between the ABJM theory and the worldvolume gauge theory of D2-branes, from identification of time-reversal invariance in these theories, and from explicit computation of Wilson loop expectation values in planar perturbation theory.

Out of these elementary Wilson loops, we can also construct composite Wilson loop operators encompassing the two product gauge groups, for example, $W_{\bf N}[C, M] \pm \overline{W}_{\overline{\bf N}}[C, M]$ or $ W_{\bf N}[C,M] \cdot \overline{W}_{\overline{\bf N}} [C, -M]$, etc. As in four-dimensional ${\cal N}=4$ super Yang-Mills theory, we expect that these Wilson loop operators constitute an important class of gauge invariant observables, providing an order parameter for various phases of the ABJM theory. In fact, even in pure Chern-Simons theory (obtainable from ABJM theory by truncating all matter fields), it was known that expectation value of Wilson loop operators yields nontrivial topological invariants~\cite{Witten:1988hf, Guadagnini:1989am}~\footnote{See \cite{Berenstein:2008dc} for an earlier discussion on Wilson loops in ABJM theory.}.

We organized this paper as follows.
In section \ref{IIA}, we collect relevant results on macroscopic IIA fundamental string in AdS$_4$, adapted from those obtained in AdS$_5$ previously. We discuss two possible configurations with different stabilizer subgroup and number of supersymmetries preserved. In section \ref{CSWL}, we formulate Wilson loop operators in ABJM theory. In subsection \ref{Def of WL}, we propose Wilson loop operators and constrain their structures by various symmetry considerations. We find from these that, up to SU(4) rotation, functional form of the Wilson loop operator is determined uniquely. Still, this leaves separate Wilson loops for U($N$) and $\overline{{\rm U}(N)}$ gauge groups, respectively. To identify relation between the two, in subsection \ref{D2 arguments}, we first recall the argument of \cite{Mukhi:2008ux,Honma:2008jd,Li:2008ya,Pang:2008hw,Honma:2008ef} relating three-dimensional super Yang-Mills theory and ABJM superconformal Chern-Simons theory~\footnote{This procedure is first proposed by Mukhi and Papageorgakis for relating (variants of) Bagger-Lambert-Gustavsson (BLG) theory\cite{Bagger:2007jr,Gustavsson:2007vu} to 3-dimensional ${\cal N}=8$ super Yang-Mills theory.}. We then identify that fundamental IIA string ending on D2-brane couples to diagonal linear combination of U($N$) and $\overline{{\rm U}(N)}$.
In section \ref{BPSWL}, we study supersymmetry condition of the Wilson loop operator and deduce that tangent field along the contour should be constant. From this, we find that unique supersymmetric Wilson loop operator is the one preserving ${1 \over 6}$ of the ${\cal N}=6$ superconformal symmetry.
In section 5, we revisit the time-reversal symmetry in ABJM theory. Based on the results of sections \ref{CSWL} and \ref{BPSWL}, we find that one combination of the elementary Wilson loops with a definite time-reversal transformation is dual to a fundamental IIA string on AdS$_4$, while orthogonal combination is mapped to zero.
In section \ref{perturbative}, we study expectation value of the Wilson loop operator to all orders in planar perturbation theory. For straight Wilson loop operator, we find that Feynman diagrams vanish identically at each loop order. For circular Wilson loop operator, we find that Feynman diagrams vanish at one loop order, nonzero at two loop order and zero again at three loop order. Remarkably, the two loop contribution consists of a part exactly the same as one-loop part of Wilson loop in ${\cal N}=4$ super Yang-Mills theory and another part exactly the same as unknotted Wilson loop in pure Chern-Simons theory. Up to three-loop orders, all Feynman diagrams involve gauge and matter kinetic terms only. Features of full-fledged ${\cal N}=6$ superconformal ABJM theory, in particular Yukawa and sextet scalar potential, begin to enter at four loops and beyond. Nevertheless, we show that the Feynman diagrams vanish identically for all odd number of loops. In other words, expectation value of the ABJM Wilson loop operator is a function of $\lambda^2$.
In section \ref{Matrix Model}, based on the results of section \ref{perturbative} and under suitable assumptions, we make a conjecture on the exact expression of circular Wilson loop expectation value in terms of a Gaussian matrix model and of unknot Wilson loop of the pure Chern-Simons theory. To match with weak and strong coupling limit results, variance of the matrix model ought to be a transcendental interpolating function of the `t Hooft coupling. Since this is different from ${\cal N}=4$ super Yang-Mills theory, we discuss issues associated with the interpolating function. Section \ref{discuss} is devoted to discussions
for future investigation. In appendix A, we collect conventions, notations and Feynman rules. In appendix B, we give details of analysis for Wilson
loops of generic contour. In appendix C, we recapitulate the one-loop vacuum polarization in ABJM theory,
obtained first in \cite{Bak:2008cp}. In appendix D, we give details for the analysis of three-loop contributions.

While writing up this paper, we noted the papers \cite{Drukker:2008zx,Chen:2008bp} posted on the arXiv archive, which have some overlap with ours.  We also found \cite{Kluson:2008wn} discuss some closely related issue.
\section{Macroscopic IIA Fundamental String in AdS$_4$} \label{IIA}
We begin with strong `t Hooft coupling regime, $\lambda \gg 1$. In this regime, by the AdS/CFT
correspondence, IIA string theory on AdS$_4 \times \mathbb{CP}^3$ is weakly coupled and provides
dual description to strongly coupled ABJM theory. As shown in \cite{Rey:1998ik,Maldacena:1998im},
correlation function of the Wilson loop operators is calculated by the on-shell action of fundamental
string whose worldsheet boundaries at the boundary of AdS space are attached to each Wilson loop operators. Following this, we shall consider a macroscopic IIA fundamental string in AdS$_4 \times \mathbb{CP}^3$ and compute expectation value of the Wilson loop operator for a straight or a circular path.

The radius of the AdS$_4$ is $L=(2{\pi^2}\lambda)^{1/4}\sqrt{\alpha'}$ as measured in unit of the IIA string tension. IIA string worldsheet configurations corresponding to straight and circular Wilson loops are exactly the same as the corresponding IIB string worldsheet configurations in AdS$_5$ background. The results are \footnote{Our convention for the relation between the IIA string coupling and rank of ABJM theory is $g_{\rm st} = 1/N$.}
\bea
&& \langle W[\mathbb{R}_t] \rangle \simeq N \nonumber \\
&& \langle W[\mathbb{S}^1] \rangle \simeq N \exp(L^2/\alpha').
\label{AdS Wilson loop result 1}
\eea
for timelike straight path $C =\mathbb{R}_t$~\cite{Rey:1998ik,Maldacena:1998im} and spacelike circular path $C = \mathbb{S}^1$~\cite{Berenstein:1998ij}, respectively. Extended to $n$ multiply stacked strings of same orientation, the ratio between the two Wilson loops is given by
\bea
{\langle W_n[\mathbb{S}^1] \rangle \over \langle W_n[\mathbb{R}_t] \rangle}
= \exp (n \sqrt{2 \pi^2 \lambda}) \, .
\eea
In IIB string theory, both string configurations are known to be supersymmetric. In section \ref{Matrix Model}, we shall try to relate these string theory results with perturbative computations in superconformal Chern-Simons theory side.

We briefly recapitulate how to get the above result. In the limit $\lambda \rightarrow \infty$, the string becomes semiclassical and sweeps out a macroscopic minimal surface in AdS-space. The metric of AdS$_4$ is expressed in Poincar\'e coordinates as
\begin{align}
 \rmd s^2=\frac{L^2}{y^2}\Big[-(\rmd x^0)^2 +(\rmd x^1)^2 + (\rmd x^2)^2 + (\rmd y)^2\Big].
\end{align}
In this coordinate system, the boundary $\mathbb{R}^{1,2}$ is located at $y=0$. We choose a macroscopic string configuration in the static gauge $x^0=\tau, y = \sigma$ and it corresponds to a timelike straight Wilson loop sitting at $x^1 = x^2 = 0$. Here, following the prescription of~\cite{Rey:1998ik,Maldacena:1998im}, we regularize the AdS-space to $y =[\epsilon, \infty]$, remove ${1 \over \epsilon}$ divergence (corresponding to self-energy) and finally lift off the regularization $\epsilon \rightarrow 0$~\footnote{Alternatively, we can prescribe renormalization scheme by adding a boundary counter-term, as in \cite{Drukker:1999zq}. The result is the same.}. The renormalized string worldsheet action is $S_{\rm ren} = 0$ and the result~\eqref{AdS Wilson loop result 1} follows.

After Wick rotation, timelike straight Wilson loop can be conformally transformed to spacelike circular Wilson loop. Let us examine this string configuration in Euclidean AdS$_4$. The metric of Euclidean AdS$_4$ is written as
\begin{align}
 \rmd s^2=\frac{L^2}{y^2}\Big[ (\rmd y)^2+(\rmd r)^2+r^2 (\rmd \theta)^2+(\rmd x)^2 \Big].
\end{align}
We choose the fundamental string configuration in the static gauge $\theta = \tau$ and $y=\sigma$, and we also take an ansatz $r = r(\sigma), \ x=0$. It corresponds to a circular Wilson loop whose center sits at $r=0$. The string worldsheet action is given by
\begin{align}
 S_{\rm ws}=\frac{1}{2\pi\alpha'}\int \sqrt{\det X^* G}
=\frac{L^2}{\alpha'}\int \rmd y \, \frac{r}{y^2}\sqrt{1+{r'}^2},
\end{align}
where $r':=\del r/\del y$. The solution with circular boundary is $r=\sqrt{1-y^2}$, and its on-shell action is written as
\begin{align}
 S_{\rm ws}=\frac{L^2}{\alpha'}\int_{\epsilon}^{1}\rmd y\frac{1}{y^2}
=\frac{L^2}{\alpha'}\left(-1+\frac{1}{\epsilon}\right).
\end{align}
Here again, we regularized the AdS-space to $y=[\epsilon, \infty]$. After removing the $1 \over \epsilon$ divergent part, we obtain the renormalized on-shell action as $S_{\rm ren}=-L^2/\alpha'$. Expectation value of the Wilson loop is $\langle W \rangle \sim \exp(-S_{\rm ren})=\exp(+L^2/\alpha')$ and the result \eqref{AdS Wilson loop result 1} follows.

We now would like to identify spacetime symmetries preserved by these classical string solutions.
Each classical string configuration wraps a suitably foliated AdS$_2$ submanifold in AdS$_4$,
so it preserves SL$(2,\mathbb{R})\times$SO(2) symmetry of the isometry SO(2,3) of AdS$_4$. If the string were sitting at a point in $\mathbb{CP}^3$, the isometry group SU(4) of $\mathbb{CP}^3$ is broken to stabilizer subgroup U(1)$\times$ SU(3). If the string were distributed over $\mathbb{CP}^1$ in $\mathbb{CP}^3$, the isometry group SU(4) is broken further to stabilizer subgroup U(1)$\times$SU(2)$\times$SU(2). Variety of other configurations are also possible, but we shall primarily focus on these two configurations. In the background AdS$_4 \times \mathbb{CP}^3$, there are 24 supercharges. They form a multiplet $({\bf 4},{\bf 6})$ of the SO(2,3)$\simeq$Sp(4,$\mathbb{R}$) and the SU(4) isometry groups. We can see that these two strings are supersymmetric by identifying supercharges that annihilate each configurations.

The first configuration turns out ${1 \over 2}$ supersymmetric. Unbroken supersymmetries ought to be organized in multiplets of the stabilizer subgroup SL$(2,\mathbb{R})\times$ SU(3). Branching rules of SO(2,3)$\times$SU(4) into SL$(2,\mathbb{R})\times$ SU(3) follows from
\begin{align}
 ({\bf 4},{\bf 6})\to ({\bf 2}+{\bf 2},{\bf 3}+\bar{\bf 3}).
\end{align}
Therefore, the minimal possibility is $({\bf 2},{\bf 3})$ of SL$(2,\mathbb{R})\times$ SU(3). Noting that ${\bf 3}$ of SU(3) is a complex representation, we deduce that the number of unbroken supercharges is either $12$ or $24$. There is no possibility that all the $24$ supercharges are preserved since the configuration does not preserve the SU(4) symmetry. So, we conclude that the string sitting at a point on $\mathbb{CP}^3$ preserves 12 of the 24 supercharges.

The second configuration is ${1 \over 6}$ supersymmetric. Branching rules of SO(2,3)$\times$SU(4)
into SL$(2, \mathbb{R})\times$SU(2)$\times$SU(2) follow from
\begin{align}
({\bf 4}, {\bf 6}) \to ({\bf 2}+ {\bf 2}, ({\bf 2}, {\bf 2}) + ({\bf 1}, {\bf 1})
+ ({\bf 1}, {\bf 1})).
\end{align}
The minimum possibility is $({\bf 2}, {\bf 1}, {\bf 1})$. Since each pair are charged oppositely under U(1), we deduce that possible number of unbroken supercharges are 4, or 16 (apart from 12 or 24 we have already analyzed). We see that a supersymmetric string distributed over $\mathbb{CP}^1$ preserves at least 4 of the 24 supercharges.

In summary, for both straight and circular string, we identified two representative supersymmetric configurations. A configuration localized in $\mathbb{CP}^3$ preserve $12$ supercharges (corresponding to ${1 \over 2}$-BPS) and SL$(2,\mathbb{R})\times$SO(2)$\times$ U(1) $\times$ SU(3) isometries. A configuration distributed over $\mathbb{CP}^1$ in $\mathbb{CP}^3$ preserves at least $4$ supercharges (corresponding to ${1 \over 6}$-BPS) and SL$(2, \mathbb{R})\times$SO(2)$\times$U(1)$\times$SU(2)$\times$SU(2) isometries.

\section{Wilson Loop: Proposal and Simple Picture} \label{CSWL}

\subsection{Wilson Loop in ${\cal N}=4$ Super Yang-Mills Theory}
We first recapitulate a few salient features of Wilson loop operator in four-dimensional ${\cal N}=4$ super Yang-Mills theory and its holographic dual, macroscopic Type IIB superstring in AdS$_5 \times \mathbb{S}^5$.
On $\mathbb{R}^{3,1}$, the Wilson loop operator for defining representation was proposed \cite{Rey:1998ik, Maldacena:1998im} to be
\bea
W_{\bf N} [C, M] = {1 \over N} {\rm Tr} \, {\cal P} \exp i \int_C \rmd \tau \Big( \dot{x}^m(\tau) A_m(x)  +  M^I(\tau) \Phi_I(x) \Big).
\label{n4symwilsonloop1}
\eea
Here, $\dot{x}^m(\tau)$ is a vector specifying $C$ in $\mathbb{R}^{3,1}$, $M^I(\tau)$ is a vector in SO(6) internal space, $A_m = A_m^a T^a \,\, (m=0,1,2,3)$ and $\Phi_I = \Phi_I^a T^a \,\, (I=1,2,3,4,5,6)$ where $T^a$s are a set of Lie algebra generators, and Tr is trace in fundamental representation. It is motivated by ten-dimensional Wilson loop operator ${1 \over N}$ Tr$ {\cal P} \exp (i \int \rmd \tau \dot{X}^M (\tau) A_M (X))$ over a path specified by $X^M(\tau)$ $(M=0, 1, \cdots, 9)$ on D9-brane worldvolume. T-dualizing to D3-brane, the gauge potential and the path are split to $(A_m(x), \Phi_I(x))$ and $(x^m(\tau), y^I(\tau))$, $(m=0,1,2,3$ and $I=1,\cdots, 6)$, respectively. We then obtain \eqref{n4symwilsonloop1}, where the vector $M^I$ is described in terms of internal coordinates as:
\bea
M^I (\tau) = \dot{y}^{\,\,I}(\tau) \, . \label{localized}
\eea
We can also motivate that this Wilson loop operator is related to Type IIB fundamental string in AdS$_5 \times \mathbb{S}^5$ by noting that $\mathbb{R}^{9,1}$ that the gauge potential $A_M(X)$ lives in is conformally equivalent to AdS$_5 \times \mathbb{S}^5$:
\bea
\rmd s^2 &=& (\rmd x^m)^2 + (\rmd y^I)^2 \nonumber \\
&=& r^2 \Big( {1 \over r^2} [(\rmd x^m)^2 + (\rmd r)^2 ] + (\rmd \Omega_5)^2\Big).
\eea
In this situation, the Wilson loop sweeps out a path in $\mathbb{R}^{9,1}$ or its conformal equivalent in AdS$_5 \times \mathbb{S}^5$.

Depending on the choice of the velocity vector $M^I(\tau)$, the Wilson loop preserves different subgroup of the SO(6) R-symmetry. If $M^I(\tau) = (0, 0, 0, 0, 0, 0)$, the Wilson loop preserves SO(6). If $M^I(\tau)$ is $\tau$-independent, the Wilson loop preserves SO(5) subgroup of SO(6) since $M^I$ can be rotated by a rigid SO(6) rotation to, say, $(|M|, 0, 0, 0, 0, 0)$. Moreover, $M^I(\tau)$ may also develop a discontinuity at some $\tau$. In holographic dual, the Wilson loop expectation value is given by a saddle-point of the string worldsheet whose boundary at AdS$_5$ infinity is prescribed by the vectors $(\dot{x}^m(\tau), M^I(\tau))$ of the Wilson loop. In general, there can be a continuous family of string worldsheets satisfying the same boundary condition, parametrized by zero-modes. In that case, each worldsheet preserves a subgroup smaller than the subgroup preserved by the corresponding Wilson loop. In order to restore the subgroup preserved by the Wilson loop, one then needs to integrate over a parameter space of the zero-modes for the string worldsheet.

One can also study the Wilson loop operators averaged over the boundary condition $M^I(\tau)$. For example,
\bea
W_{\bf N} [C, \langle M \rangle ] = {1 \over \mbox{Vol}(D(M))} \sum_{M \in D(M)} W_{\bf N}[C, M]
\label{averaged}
\eea
is an averaged Wilson loop operator in which the vector $M^I(\tau)$ is averaged to $\langle M \rangle$ over a domain $D(M)$. Each configuration of $M^I(\tau)$ preserves different subgroup of SO(6) symmetry, so the above average Wilson loop operator would retain a stabilizer subgroup common to each of $M^I(\tau)$ in $D(M)$.

\subsection{Wilson Loops in ${\cal N}=6$ Superconformal Chern-Simons Theory} \label{Def of WL}
In this subsection, paving steps parallel to the four-dimensional ${\cal N}=4$ super Yang-Mills theory, we shall construct a Wilson loop operator in the ABJM theory and find an interpretation from holographic dual side. In particular, we pay attention to features that contrast the ABJM Wilson loop operators against the Wilson loop operators in ${\cal N}=4$ super Yang-Mills theory.

Our proposal for the Wilson loop operators in the ABJM theory is as follows.
Denote coordinates of $\mathbb{R}^{1,2}$ as $x^m$ and of SU(4) internal space as $z^I, \overline{z}_I$.
With two gauge fields $A_m$ and $\overline{A}_m$ of U($N$) and $\overline{{\rm U}(N)}$ gauge groups, respectively, we can construct two types of Wilson loop operators associated with each gauge fields.
Consider the U($N$) gauge group. Our proposal of the U($N$) Wilson loop operator is
\bea
W_{\bf N} [C, M] = {1 \over N} {\rm Tr} {\cal P} \exp i \int_C \rmd \tau \, \Big(\dot{x}^m(\tau) A_m(x) + {M}_I{}^J(\tau) Y^I(x) Y^\dag_J(x) \Big). \label{ABJMwilson}
\eea
Here, $A_m = A_m^a T^a$ and $Y^I Y^\dagger_J = (Y^I Y^\dagger_J)^a T^a$, where $T^a$'s are Lie algebra generators of U($N$) gauge group. Again, the vector field $\dot{x}^m(\tau)$ specifies the path $C$ in $\mathbb{R}^{1,2}$ and ${M_I}^J(\tau)$ is a
tensor in SU(4) internal space. A choice that is a direct counterpart of (\ref{n4symwilsonloop1}) is
\be
{M_I}^J(\tau) =  \pm\left[2{\dot{\overline{z}}_I {\dot{z}^J} \over \vert \dot{{z}} \vert}  -  \delta^J_I \vert \dot{{z}} \vert\right] \, .\label{MIJ}
\ee
Since
${M_I}^J {M_J}^K = \delta_I^K \, | \dot{z}|^2 $, eigenvalues of ${M_I}^J$ are $\pm |\dot{z}|$.

We also motivate functional form of the Wilson loop from the following symmetry considerations:
\begin{list}{$\bullet$}{}

\item Wilson loop describes a trajectory of a heavy particle probe. Charge of the particle is characterized by a representations under U($N$) and $\overline{{\rm U}(N)}$ gauge groups. Mass of the particle is set by scalar fields and should carry scaling dimension 1. In (2+1) dimensions, the scalar fields $Y,Y^\dagger$ have scaling dimension 1/2. It also should transform in adjoint representation of U$(N)$. These requirements fix uniquely the requisite combination as $Y^I Y^\dagger_J$.

\item Functional form of the tensor ${M_I}^J(\tau)$ given in \eqref{MIJ} is largely determined by spacetime translational symmetry and by affine reparametrization and parity symmetries along the path $C$. Transitive motion on embedding space $\mathbb{C}^4$ is described by $z^I \rightarrow z^I + \xi^I$ for a constant $\xi^I$. The tensor is manifestly invariant under such motion since it depends only on $\dot{z}, \dot{\overline{z}}$.

\item Affine reparametrization is induced by $\tau \rightarrow \tilde{\tau}(\tau)$. The tensor ${M_I}^J$ is manifestly invariant under such motion since it transforms with Jacobian $|\rmd \tilde{\tau} / \rmd \tau|$. This cancels against the Jacobian induced by the measure $\rmd \tau$.

\end{list}
Likewise, our proposal for the Wilson loop operator of $\overline{{\rm U(}N)}$ gauge group is
\bea
\overline{W}_{\bf N} [C, M] = {1 \over N} \overline{\rm Tr} \, {\cal P} \exp i \int_C \rmd \tau \, \Big(\dot{x}^m(\tau) \overline{A}_m(x) + {M_J}^I(\tau) Y^\dag_I(x) Y^J (x) \Big),\label{ABJMwilsonbar}
\eea
where $\overline{A}_m = \overline{A}_m^a \overline{T}^a, Y^\dag_I Y^J =(Y^\dag_I Y^J)^a {T}^a$.

From ABJM theory viewpoint, various composites of these Wilson loop operators are possible (in addition to the choice of $C$ and $M$). Taking the above Wilson loop operators as building blocks, composite Wilson loops involving both gauge groups are constructible. For example, one can construct
\bea
&& W_{\bf N}[C, M] + \overline{W}_{\bf N}[C, M], \quad W_{\bf N}[C, M] + \overline{W}_{\overline{\bf N}} [C, M] \qquad
\mbox{and} \qquad ({\bf N} \leftrightarrow \overline{\bf N})  \\
&& W_{\bf N} [C, M]\cdot \overline{W}_{\bf N} [C, M], \qquad W_{\bf N} [C, M]\cdot \overline{W}_{\overline{\bf N}}[C, M]
\qquad \mbox{and} \qquad ({\bf N} \leftrightarrow \overline{\bf N})
\eea
etc. However, under suitable conditions, they turn out not independent one another. For example, at large
$N$ limit, expectation values of these composite Wilson loop operators are all equal because of large $N$ factorization property. One might have expected that the composites are further restricted if the Wilson loops are to preserve part of the ${\cal N}=6$ supersymmetry. This is not so, since supersymmetry acts on $W_{\bf N}[C,M]$ and $\overline{W}_{\bf N}[C,M]$ independently.

In comparison with ${\cal N}=4$ super Yang-Mills theory, one distinguishing feature of the ABJM theory is that there are two sets of Wilson loops, one for U($N$) gauge group and another for $\overline{{\rm U}(N)}$ gauge group. From holographic perspectives, this raises a puzzle. We expect that these Wilson loops
are mapped to a string. While there are two variety of Wilson loops in the ABJM theory, there is one and only one fundamental string in AdS$_4 \times \mathbb{CP}^3$. We first resolve this puzzle by analyzing the way
a fundamental string is coupled to a stack of D2-branes, whose worldvolume gauge theory is in turn related to the ABJM theory by moving away appropriately from conformal point.

\subsection{Fundamental String Ending on D2-Brane} \label{D2 arguments}
Consider a D2-brane and a macroscopic IIA fundamental string ending on it. From IIA supergravity field equations in the presence of the string and the D2-brane, we see that the string endpoint on the D2-brane
carries an electric charge of the worldvolume gauge field $C_m$ of the D2-brane. How is the electric charge related to charges in the ABJM theory?

Answer to this question is obtainable simply by identifying relation between the D2-brane worldvolume gauge
field $C_m$ and the two gauge fields $A_m, \overline{A}_m$ in the ABJM theory. The identification is in fact already made in \cite{Mukhi:2008ux}. By giving a nonzero vacuum expectation value to one of the bi-fundamental scalar fields in ABJM theory, one linear combination of the gauge fields becomes massive. Integrating out the massive gauge field, we are
left with orthogonal linear combination of the gauge fields. This is identified with the D2-brane worldvolume gauge field $C_m$. Relevant part of the ABJM Lagrangian is
\bea
L &=& {k \over 4\pi} \epsilon^{mnp} {\rm Tr} ( A_m \partial_n A_p +{2i \over 3} A_m A_n A_p)
 - {k \over 4 \pi}\epsilon^{mnp} \overline{\rm Tr} (\overline{A}_m \partial_n \overline{A}_p
 +{2i \over 3} \overline{A}_m \overline{A}_n \overline{A}_p) \nonumber \\
&& \quad - {\rm Tr} |\partial_m Y^I + i A_m Y^I - i Y^I \overline{A}_m |^2
- \overline{\rm Tr} |\partial_m Y^\dagger_I + i \overline{A}_m Y^\dagger_I - i Y^\dagger A_m|^2
\nonumber \\
&& \quad + {\rm Tr} A_m J^m + \overline{\rm Tr} \overline{A}_m \overline{J}^m .
\eea
The last line is to indicate how an external source with gauge currents $J_m, \overline{J}_m$ couples to the two ABJM gauge potentials.

Turn on vacuum expectation value of one of the scalar fields, say, the real part of $Y^1, Y^\dagger_1$:
\bea
\langle Y^1 \rangle = \langle Y^\dagger_1 \rangle = V \mathbb{I}_N.
\eea
We also decompose the two gauge potentials as
\bea
A_m^{(\pm)} = {1 \over 2} (A_m \pm \overline{A_m}) \, .
\eea
The corresponding field strengths are
\bea
G^{(\pm)}_{mn} = \partial_m A^{(\pm)}_n - \partial_n A^{(\pm)}_m + i [A^{(\pm)}_m, A^{(\pm)}_n] \, .
\eea
We then find that the Chern-Simons terms are reduced to
\bea
{k \over 2 \pi} \epsilon^{mnp} {\rm Tr} \,(A^{(-)}_m G^{(+)}_{np} + {2 i \over 3} A^{(-)}_m A^{(-)}_n A^{(-)}_p) + ({\rm total} \,\,\, {\rm derivative}) \, , \eea
while the kinetic terms are reduced to
\bea
4 V^2 {\rm Tr} (A^{(-)}_m)^2 + \cdots \, .
\eea
The equations of motion for $A^{(-)}_m$
\bea
A^{(-)}_m = {k \over 8 \pi V^2} {\epsilon_m}^{np} ( G^{(+)}_{np} + 2 i A^{(-)}_n A^{(-)}_p + \cdots )
\eea
can be solved perturbatively at large $k$. Collecting terms in increasing power of derivatives and redefining $g_{\rm YM}= 4 \pi V/k$, we find that the Lagrangian $L$ is reduced to
\bea
L &=& - {1 \over 2 g^2_{\rm YM}} {\rm Tr} (G^{(+)}_{mn})^2 + {\rm Tr} A^{(+)}_m (J^m + \overline{J}^m) + \cdots
\nonumber \\
&& + {4 \pi^2 \over k^2} {\cal O}({1 \over g^8_{\rm YM}} (G^{(+)})^3) + {2 \pi \over k} {1 \over g^2_{\rm YM}} \epsilon^{mnp} {\rm Tr} G^{(+)}_{mn} (J_p - \overline{J}_p).
\eea
To retain nontrivial gauge dynamics at quadratic order and suppress all higher order terms, we take
the scaling limit:
\bea
k \rightarrow \infty, \quad V \rightarrow \infty \quad {\rm and} \quad g_{\rm YM} = {4 \pi V \over k} = \mbox{fixed}.
\eea
We see that, around the vacuum given by the above expectation value, the ABJM theory is reduced to maximally supersymmetric U($N$) gauge theory of the gauge potential $A^{(+)}_m$ below the energy scale set by $g_{\rm YM}$, viz. it describes worldvolume dynamics of the D2-brane.

From the Lagrangian, we derive equations of motion for the gauge potential $A^{(+)}_m$ as
\bea
D^m G^{(+)}_{mn}= g^2_{\rm YM} (J_n + \overline{J}_n) - {2 \pi \over k}
\epsilon_{npq} D_p (J_q - \overline{J}_q) + {\cal O} (D (G^{(+)})^2).
\eea
If a fundamental string ends on the D2-brane, it acts as a source to the worldvolume gauge field $A^{(+)}_m$. In the scaling limit that reduces ABJM theory to (2+1)-dimensional super Yang-Mills theory, all but the first term drop out. This in turn implies that the string endpoint creates one unit (in unit of $g^2_{\rm YM}$) of $(J^m + \overline{J}^m)$ from ABJM currents. We also note that the non-minimal coupling of $A^{(+)}_m$ to the current $(J^m - \overline{J}^m)$ is suppressed in the above scaling limit.

In this section, we identified that $A^{(+)}_m = (A_m + \overline{A}_m)$ is the gauge field for the D2-brane worldvolume dynamics, while $A^{(-)}_m = (A_m - \overline{A}_m)$ is decoupled from the dynamics. Therefore, a fundamental string ending on D2-brane is described by the Wilson loop operator composed solely of $A^{(+)}_m$ (plus an appropriate combination of eight scalar fields). We emphasize that, under time-reversal, this Wilson loop operator transforms in the standard way. For timelike $C$, the representation ${\bf N}$ of the Wilson loop is mapped to conjugate representation $\overline{\bf N}$ but the internal tensor $M$ remains intact. For spacelike $C$, representation ${\bf N}$ remains intact but the internal tensor $M$ is mapped to conjugate
tensor $-M$.

\section{Supersymmetric Wilson loop} \label{BPSWL}
We now would like to understand under what choices of $C$ and $M_I{}^J(\tau)$ the proposed Wilson loop preserves some of the ${\cal N}=6$ superconformal symmetry. The same question was addressed previously for ${\cal N}=4$ super Yang-Mills theory \cite{Zarembo:2002an} and for the holographic dual \cite{Dymarsky:2006ve}. There, assuming that the Wilson loop sweeps a calibrated surface in $\mathbb{R}^{3,1} \times \mathbb{R}^4$, it was found that the Wilson loop preserving $1/2$ of the ${\cal N}=4$ superconformal symmetry ought to lie in $\mathbb{R}^{3,1}$ on either a timelike straight path or a spacelike circular path.
Here, we shall check if the same choice of $C$ of the ABJM Wilson loop operators is supersymmetric.
More general choice of the contour $C$ will be discussed later in this section.

Begin with the ABJM Wilson loop over a timelike straight path. By a Lorentz boost, we can always bring
the path to $x^m(\tau) = (\tau, 0, 0)$, so $\dot{x}^m = (1,0,0)$. We first focus on the U($N$) Wilson loop operator:
\bea
W_{\bf N}[C, M] = {1 \over N} \mbox{Tr}\,  {\cal P} \, \exp \Big( i \int_{-\infty}^{\infty} \rmd \tau \,  (A_0+M_{I}{}^{J}Y^{I}\Yb_{J})\Big) \, . \label{st-line}
\eea
As in \cite{Zarembo:2002an, Dymarsky:2006ve}, we take the ansatz that $M_{I}{}^{J}$ is a $\tau$-independent, constant tensor.

The ${\cal N}=6$ Poincar\'e supersymmetry transformations for the gauge and scalar fields are~\cite{Gaiotto:2008cg,Hosomichi:2008jb,Terashima:2008sy}
\bea
&& \delta Y^I=2 i \xi^{IJ}\psib_{J},\qquad \delta \Yb_{I}=2i\xi_{IJ}\Psi^{J},\\
&& \delta A_{m}=2\xi_{IJ}\gamma_{m}Y^{I}\psi^{J}+2\psib_{J}\Yb_{I}\gamma_{m}
\xi^{IJ},
\eea
where $\xi^{IJ},\ \xi_{IJ}$ are supersymmetry parameters satisfying the following relations:
\bea
&& \xi_{IJ}=-\xi_{JI},\qquad \xi^{IJ}:=\frac{1}{2}\epsilon^{IJKL}\xi_{KL},\qquad
(\xi_{IJ})^{*}=\xi^{IJ}.\label{parameter relation}
\eea
Consider a point $\tau$ along the contour $C$. The supersymmetry variation of the integrand in the exponent of \eqref{st-line} becomes
\bea
\delta \left(A_0+M_{I}{}^{J}Y^{I}\Yb_{J}\right)
=2\left(\xi_{IJ}\gamma_0+ i M_{I}{}^{K}\xi_{KJ}\right)Y^{I}\psi^{J}
-2\left(\xi^{IJ}\gamma_0- i M_{K}{}^{I}\xi^{KJ}\right)\psib_{J}\Yb_I \, .
\eea
In order to be supersymmetric, the following two equations must be satisfied for some of the supersymmetry parameters:
\bea
&& \xi_{IJ}\gamma_0+i M_{I}{}^{K}\xi_{KJ}=0,\qquad \xi^{IJ}\gamma_0-i M_{K}{}^{I}\xi^{KJ}=0.\label{var}
\eea

By unitary transformation, diagonalize the constant Hermitian matrix ${M_I}^J$ as
\bea
 M =U \Lambda U^{-1},\qquad \mbox{where} \qquad \Lambda=\mathrm{diag}(\lambda_1,\lambda_2,\lambda_3,\lambda_4).
\eea
In this frame, the supersymmetry condition \eqref{var} reads
\bea
\xi_{IJ}\gamma_0+i \lambda_{I}\xi_{IJ}=0; \qquad \quad \xi^{IJ}\gamma_0- i \lambda_I \xi^{IJ}=0 \, \qquad (\text{no summation over $I$}).
\label{SUSY condition}
\eea
We see that each eigenvalues $\lambda_{I}$ must take values $\pm 1$ in order to satisfy the conditions \eqref{SUSY condition}. If one of the eigenvalues, say $\lambda_{1}$, is not $\pm 1$,  since the eigenvalues of $\gamma_{0}$ are $\pm i$, \eqref{SUSY condition} implies $\xi^{1J}=0, \xi_{1J}=0,\ (J=2,3,4)$. In this case, the second relation of \eqref{parameter relation} reads $\xi^{IJ}=\xi_{IJ}=0$ for $I,J=2,3,4$ as well and no supersymmetry is preserved.

Modulo overall sign and permutations of the eigenvalues, there are three possible combinations. We examine each of them separately.
\begin{itemize}
\item \underline{$M=\mathrm{diag}(+1, +1, +1, +1)$}: \hfill\break
This configuration preserves full SU(4) symmetry. The supersymmetry conditions \eqref{SUSY condition} now read
\begin{align}
\xi_{IJ}\gamma_0+ i \xi_{IJ}=0,\qquad \xi^{IJ}\gamma_0- i\xi^{IJ}=0.
\end{align}
These two equations cannot be satisfied simultaneously because of the reality condition \eqref{parameter relation}. So, there is no supersymmetric Wilson loop with unbroken SU(4) symmetry.
The same conclusion holds for $M= \mathrm{diag}(-1,-1,-1,-1)$.

\item \underline{$M=\mathrm{diag}(-1,+1,+1,+1)$}: \hfill\break
This configuration breaks SU(4) to SU(3)$\times$U(1).
From the supersymmetry condition \eqref{SUSY condition} for $(I,J)=(1,J)$ and $(2,J)$ 
and the first relation of \eqref{parameter relation}, it follows that $\xi^{1J}=\xi_{1J}=0$. 
This and the second relation of \eqref{parameter relation} imply that $\xi^{IJ}=\xi_{IJ}=0$ 
for all $I,J=1,2,3,4$.
Again, there is no supersymmetric Wilson loop with unbroken SU(3)$\times$U(1) symmetry. 
The same conclusion holds for $M=\mathrm{diag}(+1,-1,-1,-1)$.

\item \underline{$M=\mathrm{diag}(-1,-1,+1,+1)$}: \hfill\break
This configuration breaks SU(4) to SU(2)$\times$SU(2)$\times$U(1).
In this case, supersymmetry parameters $\xi^{12}$ and $\xi^{34}$ satisfying the projection conditions:
\begin{align}
\xi^{12}\gamma_0+ i \xi^{12}=0,\qquad \xi^{34}\gamma_0 - i  \xi^{34}=0.
\end{align}
exists. Other components of $\xi^{IJ}$ should vanish. We thus find that this Wilson loop preserves $2$ 
real supercharges. Since conformal supersymmetry transformations of $A_m, Y^I, Y^\dagger_I$ are obtainable 
from Poincar\'e supersymmetry by the substitution $\xi^{IJ} \rightarrow \gamma_m x^m \widetilde{\xi}^{IJ} $, 
we also find that this Wilson loop preserves $2$ real conformal supercharges.
We conclude that this Wilson loop preserves ${1 \over 6}$ of the ${\cal N}=6$ superconformal symmetry.
\end{itemize}
In summary, the supersymmetric Wilson loop in ABJM theory is unique: it has the tensor ${M_I}^J$ which
has maximal rank $M=\mathrm{diag}(-1,-1,+1,+1)$, preserves SU(2)$\times$SU(2)$\times$U(1) symmetry
of SU(4), and corresponds to a ${1 \over 6}$-BPS configuration of the ${\cal N}=6$ superconformal symmetry
\footnote{There are other supersymmetric configurations. For example, a ${1 \over 3}$-BPS configuration is 
obtainable by $\dot{x}^m=0$ and ${M_I}^J = \delta_I^1 \delta^J_4$. However, since $\dot{x}^m=0$, this 
configuration is actually a generating functional of all ${1 \over 3}$-BPS {\sl local} operators. 
A direct counterpart in ${\cal N}=4$ super Yang-Mills theory is the $\dot{x}^m=0$ and 
$M^I = (0,0,0,0,1, i)$ configuration. Again, with $\dot{x}^m = 0$, this Wilson loop is a generating 
functional of ${1 \over 2}$-BPS {\sl local} operators \cite{mine} (see also \cite{miwa, fujita}). }.

Actually, the Wilson loop operator \eqref{ABJMwilson} is closely related to the Wilson loop considered in \cite{Gaiotto:2007qi} in ${\cal N}=2$ superconformal Chern-Simons theory. The ${1 \over 6}$-BPS configuration we found above is the same as the ${1 \over 2}$-BPS configuration of the ${\cal N}=2$ superconformal symmetry: for a straight timelike path, both preserves two Poincar\'e supersymmetries and two conformal supersymmetries. So, features we find in this paper ought to hold to various ${\cal N}=2$ superconformal Chern-Simons theories.

Notice that the tensor ${M_I}^J$ of the ${1 \over 6}$-BPS configuration has the properties ($n=$ positive integer)
\bea
\mbox{Tr} M^{2n-1}=0 \qquad \mbox{and} \qquad \mbox{Tr} M^{2n} = 4.
\eea
Though trivial looking, these properties will play a crucial role when we evaluate in the next section the Wilson loop expectation value explicitly in planar perturbation theory.

We can also generalize the supersymmetric Wilson loops to a general contour $C$ specified by tangent vector $\dot{x}^m(\tau)$. The supersymmetry condition now reads
\bea
&& \xi_{IJ}\gamma_m \dot{x}^m(\tau)+M_{I}{}^{K}(\tau)i\xi_{KJ}=0,\qquad
\xi^{IJ}\gamma_m \dot{x}^m(\tau)-M_{K}{}^{I}(\tau)i\xi^{KJ}=0 \, . \label{var'}
\eea
We assume that $C$ is smooth, implying that $\dot{x}^m(\tau)$ is a smooth function of $\tau$. We also set $|\dot{x}(\tau)|=1$ using the reparametrization invariance.
The important point is that (\ref{var'}) ought to satisfy the supersymmetry conditions at each $\tau$.
Without loss of generality, we assume at $\tau=0$ that $M(0)=\mathrm{diag}(-1,-1,+1,+1)$ and
the only non-zero components of $\xi_{IJ}$ are
$\xi_{12}$ and $\xi_{34}$: these are the eigenstates of $\gamma_m \dot{x}^m(0)$ with eigenvalue $+i$ and $-i$, respectively. It is then possible to show that (\ref{var'}) allows only a constant $M(\tau)$ and $x^m(\tau)$. The details of the proof of this statement is given in Appendix \ref{proof}. In plain words, tangent vector $\dot{x}^m$ along the contour $C$ should remain constant. We conclude that the Wilson loop is supersymmetric only if $C$ is a straight line. The circular Wilson loop, which is a conformal transformation of this supersymmetric Wilson loop, is annihilated not by the Poincar\'e supercharges, but by linear combinations of the Poincar\'e supercharges and the conformal supercharges. The conformal transformation on $\mathbb{R}^{1,2}$ cannot affect ${M_I}^J$. So, $M=\mathrm{diag}(-1,-1,+1,+1)$ is also the tensor relevant for the circular supersymmetric Wilson loops.

Still, the above result poses a puzzle. We argued that the Wilson loops proposed are unique in the sense that the supersymmetry considerations fix its structure completely. We also found that these Wilson loops preserve ${1 \over 6}$ of the ${\cal N}=6$ supersymmetry, but no more. On the other hand, the macroscopic IIA fundamental string preserves ${1 \over 2}$ of the ${\cal N}=6$ supersymmetry. At present, we do not have a
satisfactory resolution. We expect that the supersymmetric Wilson loop corresponds to a string worldsheet whose location on $\mathbb{CP}^3$ is averaged over, perhaps, in a manner similar to the prescription \eqref{averaged}.
An encouraging observation is that the R-symmetry preserved by the Wilson loop is the same as the isometry preserved by the
string smeared over $\mathbb{CP}^1$ in $\mathbb{CP}^3$, and the number of preserved supercharges also match. This also fits to the observation that $M=\mathrm{diag}(-1,-1,+1,+1)$ above cannot be written as (\ref{MIJ}) for any choice of $z^I(\tau)$ since the trace of (\ref{MIJ}) does not vanish.

\section{Consideration of Time-Reversal Symmetry}
Though it involves Chern-Simons interactions, the ABJM theory is invariant under (suitably generalized) time-reversal transformations. This also fits well with the observation in section 3.2 that, by vacuum expectation value of scalar fields, the ABJM theory is continuously connected to the worldvolume gauge theory of multiple D2-branes. The latter theory is invariant under parity and time-reversal transformations. In section 3.2, we also identified $A^{(+)}_m = {1 \over 2} (A_m + \overline{A}_m)$ as the right combination of the ABJM gauge potentials that couples to the current $(J^m + \overline{J}^m)$ of the string endpoint on D2-brane. We shall now combine this observation and time-reversal transformation properties to identify $\langle {\cal W}_{\bf N}[C, M] \rangle$, where
\bea
{\cal W}_{\bf N}[C, M] \Big\vert_{\rm timelike} := {1 \over 2} \Big( W_{\bf N}[C, M] + \overline{W}_{\bf N} [C, M] \Big)_{\rm timelike}  \, ,
\label{Teven} \eea
as the timelike Wilson loop dual to the fundamental IIA string. We shall now show that \eqref{Teven} transforms under the time-reversal precisely the same as the D2-brane worldvolume gauge potential that couple to the fundamental string. Moreover, since the other orthogonal combination $A^{(-)}_m = {1 \over 2} (A_m - \overline{A}_m)$ is not present in the worldvolume gauge theory of multiple D2-branes, we are led to identify that expectation value of Wilson loops for the other combination vanishes identically:
\bea
\Big\langle W_{\bf N} [C, M] - \overline{W}_{\bf N} [C, M] \Big\rangle_{\rm timelike} = 0.
\label{Todd} \eea

Consider a timelike Wilson loop $W_{\bf N}[C, M]$ in $\mathbb{R}^{1,2}$. We take its path $C$ along the time direction, $\dot{x}^m = (1,0,0)$. By definition,
\begin{eqnarray}
W_{\bf N} [C, M]
&=& {1 \over N} \mbox{Tr} \, {\cal P} \, \exp i\int_C \rmd \tau(\Phi(\tau)) \nonumber \\
&:=& {1 \over 2} \sum_{n=0}^\infty i^n\int_{\tau_1>\cdots>\tau_n}\mbox{Tr}\langle \Phi(\tau_1)\cdots \Phi(\tau_n)\rangle,
\end{eqnarray}
where $\Phi$ denotes exponent of the Wilson loop:
\begin{equation}
\Phi(\tau) = T^a \Big[A^a_0(x)+M_I{}^J (Y^I Y^\dag_J)^a (x)\Big]_{x = x(\tau)}.
\end{equation}

Under the time-reversal transformation, $x^m = (x^0, x^1, x^2) \rightarrow \tilde{x}^m = (-x^0, x^1, x^2)$. In the ABJM theory, this is adjoined with $\mathbb{Z}_2$ involution that exchanges the two gauge groups
U($N$) and $\overline{{\rm U}(N)}$. The resulting generalized time-reversal transformation $T$ then acts on relevant fields as
\bea
T \, \Big( A^a_0(x), \overline{A}^a_0(x), Y^I(x), Y^\dagger_I(x) \Big) \, T^{-1} = \Big( \overline{A}^a_0(\tilde{x}), A^a_0 (\tilde{x}), Y^\dagger_I(\tilde{x}), Y^I(\tilde{x}) \Big) \, .
\eea
Being anti-linear, $T$ also acts as
\bea
T(i)T^{-1} = - i.
\label{antiunitary}
\eea
Moreover, since the path $C$ is timelike, $T$ also reverses ordering of the path. To bring the path
ordering back, we take transpose of products of $\Phi(\tau)$s inside trace. Together with minus sign from time reversal, the generators $T^a$ are mapped to $-(T^a)^{\rm T} = \overline{T}^a$. These are the generators for the complex conjugate representation. Thus, the exponent of the timelike Wilson loop transforms as
\begin{equation}
T \, \Phi(\tau) T^{-1} = \overline{\Phi}(-\tau),
\end{equation}
where
\begin{equation}
\overline{\Phi}(\tau) = \overline{T}^a[\overline{A}^a_0(\tau)+M^I{}_J (Y^\dag_IY^J)^a(\tau)].
\end{equation}
We see that the time-reversal $T$ acts on the Wilson loop $W_{\bf N} [C, M]$ as
\bea
T\,  \Big( W_{\bf N}[C, M] \Big) \, T^{-1} = \overline{W}_{\overline{\bf N}} [C, M]; \qquad
T\, \Big(\overline{W}_{\bf N} [C, M] \Big) \, T^{-1} = W_{\overline{\bf N}} [C, M].
\label{T-transform}\eea
Notice, however, that $T$ does not change the path $C$ and the internal tensor ${M_I}^J$.

With \eqref{T-transform}, we identify that \eqref{Teven} is the linear combination of elementary Wilson loops that transform under the generalized time-reversal transformation $T$:
\bea
T: \qquad {\cal W}_{\bf N}[C, M] \Big\vert_{\rm timelike} \quad \longrightarrow \quad {\cal W}_{\overline{\bf N}}[C, M] \Big\vert_{\rm timelike}.
\eea
This is precisely how the Wilson loop operator on D2-brane worldvolume behaves (as derived at the end of section 3): under the time-reversal,
the Wilson loop of $A^{(+)}_m$ gauge field in the representation ${\bf N}$ transforms to the Wilson loop in representation $\overline{\bf N}$.
Moreover, by expanding the Wilson loops, we see that the contour $C$ couples to $(A^a_m + \overline{A}^a_m) {T}^a$. In section 3.2, we identified this combination with the gauge field $A^{(+)}_m$ on the D2-brane worldvolume that couples to the fundamental string.  As such, the path $C$ is identifiable with trajectory of the fundamental string endpoint at the boundary of AdS$_4$.
On the other hand, we see that the linear combination of Wilson loops in \eqref{Todd} represent $(A^a_m - \overline{A}^a_m) T^a$ along the contour $C$. This is the gauge field $A^{(-)}_m$ that was lifted up nondynamical out of the D2-brane worldvolume dynamics. We thus conclude that vacuum expectation value \eqref{Todd} ought to
vanish identically.
\vspace{5mm}

Consider next the Wilson loop with path $C$ a spacelike circle in $\mathbb{R}^{1,2}$. By conformal transformation, we can put radius of the circle to 1 and parametrize $C$ by $\dot{x}^m(s) = (0, \cos s, \sin s)$, $s = [0, 2 \pi]$. In this case, the exponent $\Phi(s)$ is given by
\begin{equation}
\Phi(s) = T^a [\dot{x}^i A^a_i(x) + M_I{}^J (Y^IY^\dag_J)^a(x)]_{x = x(s)}.
\end{equation}
Now, under $T$, the spatial components of the gauge potential are transformed by
\bea
T \, \Big( A^a_i(x), \overline{A}^a_i(x) \Big) \, T^{-1} = \Big( - \overline{A}^a_i(\tilde{x}), - A^a_i (\tilde{x}) \Big) \, .
\eea
Since the path $C$ is spacelike, under $T$, its path ordering and hence the Lie algebra generators $T^a$s remain unchanged. Thus, with the anti-linearity \eqref{antiunitary} taken into account, the exponent of the spacelike circular Wilson loop transforms as
\begin{equation}
T \, \Phi(s) T^{-1} = \overline{\Phi}(s),
\end{equation}
where
\begin{equation}
\overline{\Phi}(s) = {T}^a[\dot{x}^i(s)\overline{A}^a_i(s) - M^I{}_J (Y^\dag_IY^J)^a(s)].
\end{equation}
We see that the time-reversal $T$ acts on the spacelike Wilson loop $W_N [C, M]$ as
\bea
T\,  \Big( W_{\bf N}[C, M] \Big) \, T^{-1} = \overline{W}_{\bf N} [C, - M]; \qquad
T\, \Big(\overline{W}_{\bf N} [C, M] \Big) \, T^{-1} = W_{\bf N} [C, - M].
\label{T-transform2}\eea
Notice that $T$ now flips sign of the internal tensor ${M_I}^J$.

With the transformation \eqref{T-transform2}, we now identify for spacelike circular Wilson loops that
\bea
{\cal W}_{\bf N} [C, M] \Big\vert_{\rm spacelike} := {1 \over 2} \Big( W_{\bf N} [C, M] + \overline{W}_{\bf N} [C, M] \Big)_{\rm spacelike}
\eea
is the linear combination that transforms requisitely under the generalized time-reversal transformation $T$:
under time-reversal, spacelike Wilson loop operator on the D2-brane worldvolume transforms as
\bea
T: \qquad {\cal W}_{\bf N}[C, M] \quad \longrightarrow \quad {\cal W}_{{\bf N}} [C, -M].
\eea
By expanding the Wilson loops, we again find that the spacelike path $C$ couples to the correct linear combination of gauge potentials, $(A^a_m + \overline{A}^a_m) T^a$. On the other hand, by a reasoning parallel to the timelike Wilson loops, we learn that
\bea
\Big\langle W_{\bf N} [C, M] - \overline{W}_{\bf N} [C, M] \Big\rangle_{\rm spacelike} = 0.
\eea
%

\section{Perturbative Computation} \label{perturbative}
In this section, we compute expectation value of the elementary Wilson loop operator
$\langle W_{\bf N}[C, M] \rangle$
in planar perturbation theory. Prompted by the conclusions of previous sections, we choose the contour $C$
either a timelike line or
a spacelike circle. For this purpose, we expand the Wilson loop expectation value in powers of the phase factor.
Start with the definition of the Wilson loop operator in Lorentzian spacetime $\mathbb{R}^{1,2}$:
\begin{eqnarray}
\langle W_{\bf N}[C, M] \rangle
 &=& {1 \over N} \sum_{n=0}^\infty i^{\, n} \int_{-\infty}^{+\infty}\rmd \tau_1 \int_{-\infty}^{\tau_1} \cdots\int_{-\infty}^{\tau_{n-1}} \rmd \tau_n\   \\
 & & \hspace*{0.5cm} \Bigl\langle\mbox{Tr}\left[\{A_0(\tau_1)+M_I{}^JY^IY^\dag_J(\tau_1)\}\cdots\{A_0(\tau_n)+M_I{}^JY^IY^\dag_J
 (\tau_n)\}\right]
     \Bigr\rangle .
   \nonumber
\end{eqnarray}
We shall perform perturbative evaluation in Euclidean spacetime $\mathbb{R}^3$. In this case, the exponent of the Wilson loop is changed to
\bea
A_0(\tau)\rmd \tau \to A_m(x(\tau))\dot{x}^m(\tau)\rmd\tau, \qquad \qquad  M_I{}^J \to i \,
M_I{}^J. \label{replacement}
\eea
Computations of $\langle W_{\overline{\bf N}} [C, M] \rangle, \langle \overline{W}_{\bf N}[C, M] \rangle$ or $\langle \overline{W}_{\overline{\bf N}}[C, M]\rangle$ etc. proceed exactly the same.

To evaluate Feynman diagrams in momentum space \footnote{Evaluation of Feynman diagrams in coordinate space are completely parallel and equally efficient.}, we rewrite the above expansion of the Wilson loop as follows:
\begin{eqnarray}
\langle W_{\bf N}[C, M] \rangle
 &=& {1 \over N} \sum_{n=0}^\infty  i^{\, n} \int_{-\infty}^{+\infty} \rmd \tau_1\int_{-\infty}^{\tau_1}\cdots\int_{-\infty}^{\tau_{n-1}}\rmd \tau_n\
     \int_{p_1}\cdots\int_{p_n}\ e^{i(p_1^0t_1+\cdots+p_n^0t_n)} \nonumber \\
 & & \Bigl\langle \mbox{Tr}\left[\{A_0(p_1)+YY^\dag(p_1)\}\cdots
     \{A_0(p_n)+YY^\dag(p_n)\}\right]\Bigr\rangle ,
\end{eqnarray}
Action, Feynman rules and conventions of the ABJM theory needed for perturbation theory are summarized in Appendix \ref{notation}.

\begin{figure}
\epsfxsize=2.2in
\begin{center}
\includegraphics[scale=0.7]{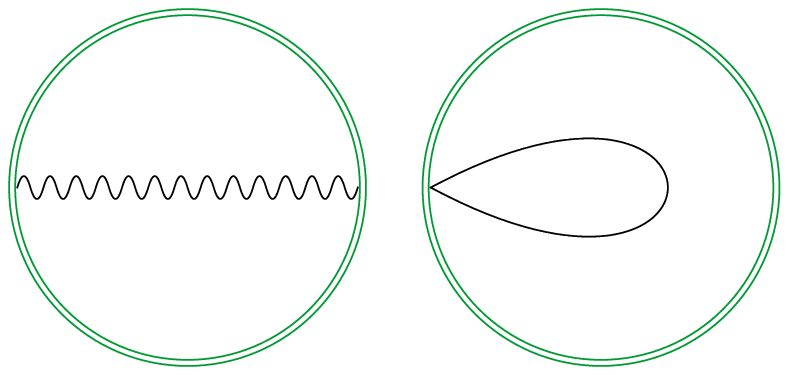}
\end{center}
\caption{\sl The Feynman diagrams contributing at order $\lambda^1$.}
\label{oneloop}
\end{figure}

\begin{figure}
\epsfxsize=2.2in
\begin{center}
\includegraphics[scale=0.7]{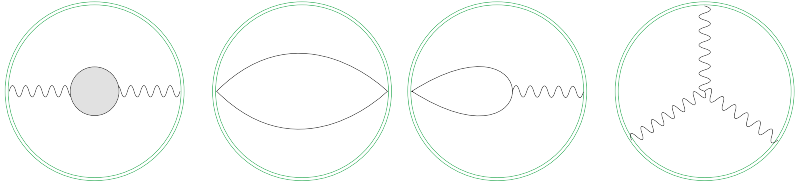}
\end{center}
\caption{\sl The Feynman diagrams contributing at order $\lambda^2$.}
\label{twoloop}
\end{figure}

\vspace{5mm}

Planar perturbative contribution to $W_{\bf N}[C, M]$ is organized in powers of the 't Hooft coupling $\lambda$ in (\ref{tHooft}) as
\begin{equation}
\langle W_{\bf N}[C, M] \rangle = \sum_{n=0}^\infty W_n[C]\lambda^n,
\end{equation}
with $W_0[C]=1$. We shall evaluate $W_1, W_2, W_3$ explicitly, and then establish vanishing theorem that $W_n$ vanishes for odd $n$ to all orders in planar perturbation theory.

\vspace{5mm}

\subsection{$W_1[C]$}

\vspace{5mm}

It is straightforward to check that all one-loop diagrams contributing to $W_1[C]$ vanish identically.
The relevant diagrams are depicted in fig.~\ref{oneloop}\footnote{
For $C$ a timelike line, the relevant Feynman diagrams are obtained by cutting the contour $C$ in the figures at a point and identifying the two ends with $\tau = \pm \infty$. Different choices of the point generate all combinatorially different diagrams. }.

The first diagram vanishes by itself. For $C$ the timelike line, the diagram is proportional to $\epsilon^{00m}$ and vanishes trivially. For $C$ the spacelike circle, the diagram is proportional to
\begin{equation}
\dot{x}^m(\tau_1)\dot{x}^n(\tau_2)\langle A_m(\tau_1)A_n(\tau_2)\rangle
\quad \propto \quad \dot{x}^m(\tau_1)\dot{x}^n(\tau_2)\epsilon_{mnk}
\frac{(x(\tau_1)-x(\tau_2))^k}{|x(\tau_1)-x(\tau_2)|^3} \, .
\end{equation}
As the vector $\dot{x}^m(\tau)$ is contained within $\mathbb{R}^2$, this again vanishes identically.

The second diagram in fig.~\ref{oneloop} also vanishes by itself. For $C$ both the timelike line and the spacelike circle, the diagram is proportional to Tr$M$. In the previous section, we found that supersymmetry of the Wilson loop imposes Tr$M$ to vanish. It is also worth mentioning that the operator $M_I{}^JY^IY^\dag_J$ is automatically normally-ordered if Tr$M=0$.

\vspace{5mm}

\subsection{$W_2[C]$ \label{lambda2}}

\vspace{5mm}

The two-loop Feynman diagrams contributing to $W_2[C]$ are summarized in fig. \ref{twoloop}.

\begin{figure}
\epsfxsize=2.2in
\begin{center}
\includegraphics[scale=0.7]{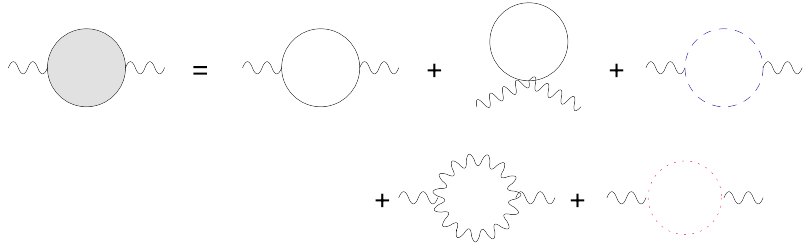}
\end{center}
\caption{\sl One loop photon self energy diagrams from bosons, Faddeev-Popov ghosts, gauge bosons, fermions, respectively. Contributions of boson tadpole vanishes identically. Contributions of Faddeev-Popov ghosts and gauge bosons cancel each other.}
\label{pse1}
\end{figure}

Begin with $C$ the timelike line.
The first diagram in fig. \ref{twoloop} involves the vacuum polarization tensor $\Pi^{mn}(p)$ depicted in fig. \ref{pse1}. At one loop, it gives parity and time-reversal invariant contribution:
\begin{equation}
\frac{N}{2|p|}(p^m p^n-\eta^{mn}p^2) \, ,
\end{equation}
The derivation is recapitulated from \cite{Bak:2008cp} (see also \cite{Gaiotto:2007qi})
in Appendix \ref{polarization}. Utilizing this, the first diagram in fig. \ref{twoloop} yields
\begin{equation}
{ i^2 \over N} 4\pi^2 \lambda^2\frac{\epsilon_{0l m}p^l}{p^2}\frac{N}{2|p|}(p^m p^n-\eta^{mn}p^2)
\frac{\epsilon_{0k n}p^k}{p^2}
\quad = \quad 2\pi^2 \lambda^2\frac1{|p|}\left[ 1+\frac{(p^0)^2}{p^2} \right].
  \label{gaugeprop}
\end{equation}
The first term in (\ref{gaugeprop}) is canceled by the second diagram in fig. \ref{twoloop}.
In computing the second diagram in fig. \ref{twoloop}), we used the supersymmetry condition Tr$M^2=4$, which counts the number of matter flavors in ABJM theory.
However, this should not be taken as a restriction on the matter content of the theory.
The first diagram in fig. \ref{twoloop} is also proportional to the number of matter flavors, so the cancelation persists for any number of matter flavors.
The non-covariant term in (\ref{gaugeprop}) vanishes since the contour integral generates $\delta(p^0)$.

The remaining diagrams in fig. \ref{twoloop} vanish separately. The third diagram vanishes since it involves $\mbox{Tr}M=0$. The fourth diagram vanishes since it is proportional to $\epsilon^{00m}$.

\vspace{5mm}

Consider next $C$ the spacelike circle. In this case, a remarkable structure emerges.
Recall that the one-loop correction to gluon propagator is parity and time-reversal invariant. In Feynman gauge, it takes the form \cite{Gaiotto:2007qi}
\begin{equation}
\langle A^a_m(x)A^b_n(y) \rangle =
\frac{2N}{k^2}\delta^{ab}\left[ \frac{\eta_{mn}}{(x-y)^2}-\frac12\partial_m\partial_n\log (x-y)^2 \right].
   \label{glue}
\end{equation}
Treating this as gauge boson skeleton propagator, the first diagram in fig. \ref{twoloop} is obtained.
Likewise, the second diagram in fig. \ref{twoloop} is obtained by treating the one-loop as scalar-pair skeleton propagator:
\begin{equation}
\langle {M_I}^J Y^I Y^\dagger_J(x) {M_K}^L Y^K Y^\dagger_L(y) \rangle = N \, \mbox{Tr}M^2\left[ \frac{2\pi}{k}\frac1{4\pi |x - y\,|} \right]^2.
\end{equation}
Taking account of Tr $M^2 = 4$ and (\ref{replacement}), these skeleton propagators put the contribution
from the first two diagrams in fig. \ref{twoloop} to \footnote{This formula holds for any contour $C$. For the timelike line, the integrand vanishes identically, reproducing the result obtained below (\ref{gaugeprop}).}
\begin{equation}
{1 \over N} \, N\lambda^2\int_{\tau_1>\tau_2}\frac{-\dot{x}(\tau_1)\cdot \dot{x}(\tau_2)+|\dot{x}(\tau_1)||\dot{x}(\tau_2)|}
{(x(\tau_1)-x(\tau_2))^2} = {1 \over 2^2} (2\pi)^2 \, \lambda^2.
  \label{importantresult}
\end{equation}
Here, we used the fact that the second term in (\ref{glue}) vanishes after the contour integration.
Remarkably, this {\sl two-loop} contribution has exactly the same functional form in configuration space as the {\sl one-loop} contribution to supersymmetric Wilson loops in four-dimensional ${\cal N}=4$ super Yang-Mills theory \cite{Erickson:2000af}. In the latter theory, assuming that all vertex-type diagrams do not contribute, the circular Wilson loop expectation value was mapped to a zero-dimensional Gaussian matrix model. Strong `t Hooft coupling limit of the Gaussian matrix model matched well with minimal surface result in string theory side. In the next section, we will take the same assumption on vertex-type diagrams, utilize the above observation on skeleton propagators, and propose a conjecture concerning circular Wilson loop in ABJM theory in terms of a Gaussian matrix model.

The fourth diagram in fig. \ref{twoloop} is also encountered in the context of pure Chern-Simons theory, and its value is well-known
\cite{Guadagnini:1989am}.
We obtain
\begin{equation}
{i^2 \over N} \, \frac{N\lambda^2}{16\pi}\int_{\tau_1>\tau_2>\tau_3}\dot{x}(\tau_1)^l \dot{x}(\tau_2)^m \dot{x}(\tau_3)^n\epsilon^{abc}
   \epsilon_{lai}\epsilon_{mbj}\epsilon_{nck}I^{ijk} = -\frac{\pi^2 \lambda^2}{6},
\end{equation}
where
\begin{equation}
I^{ijk} =
\int \rmd^3x\frac{(x-x(\tau_1))^i(x-x(\tau_2))^j(x-x(\tau_3))^k}{|x-x(\tau_1)|^3|x-x(\tau_2)|^3|x-x(\tau_3)|^3}.
\end{equation}

\vspace{5mm}

We summarize the computations so far. For the timelike line,
\begin{equation}
\langle W_{\bf N}[C, M] \rangle = 1 + {\cal O}(\lambda^3).
\end{equation}
For the spacelike circle,
\bea
\langle W_{\bf N}[C, M] \rangle = \Big( 1 + \pi^2\lambda^2 + \cdots \Big) \Big( 1 -\frac{\pi^2\lambda^2}{6} + \cdots \Big) + {\cal O}(\lambda^3).
\label{spacelikeWilson2loop}
\eea
The first part is identical to the circular Wilson loop in four-dimensional ${\cal N}=4$ super Yang-Mills theory, while the second part is identical to the unknotted Wilson loop in pure Chern-Simons theory.

\begin{figure}
\begin{center}
\includegraphics[scale=0.7]{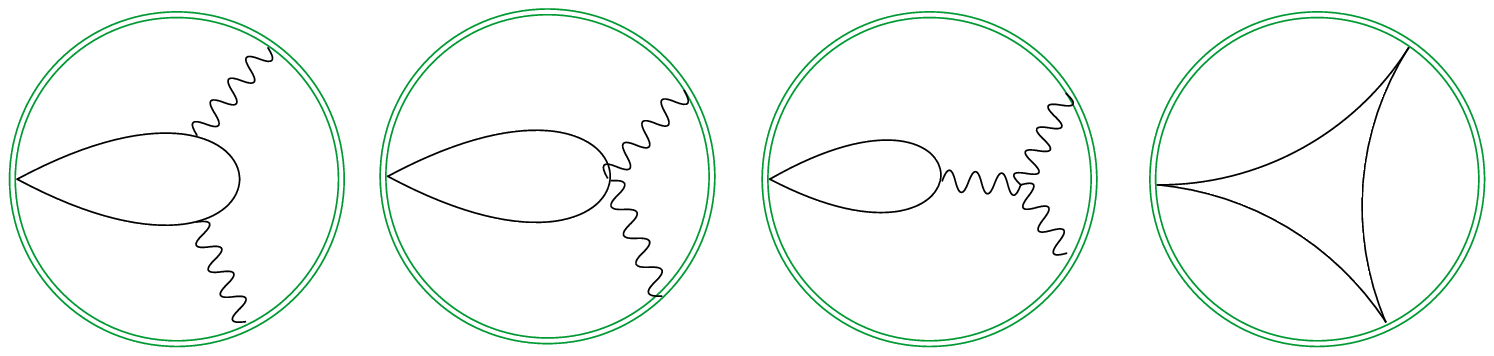}\\
\includegraphics[scale=0.8]{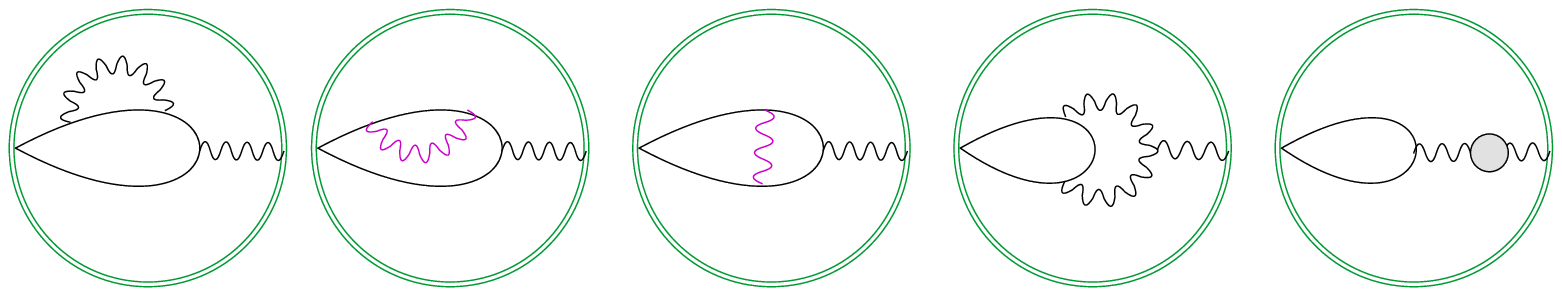}
\end{center}
\caption{\sl The diagrams of order $\lambda^3$ which vanish by themselves.}
\label{3-3gss-sss}
\end{figure}

\vspace{5mm}

\subsection{$W_3[C]$}

\vspace{5mm}

We next analyze three-loop diagrams contributing to $W_3[C]$ and show that they all vanish.

Consider the timelike line first.
All Feynman diagrams listed in fig. \ref{3-3gss-sss} vanish identically because they are proportional to the supersymmetry conditions
\begin{equation}
\mbox{Tr}M = 0 \qquad \mbox{and} \qquad \mbox{Tr}M^3 = 0,
\end{equation}
respectively.

\begin{figure}
\epsfxsize=2.2in
\begin{center}
\includegraphics[scale=0.7]{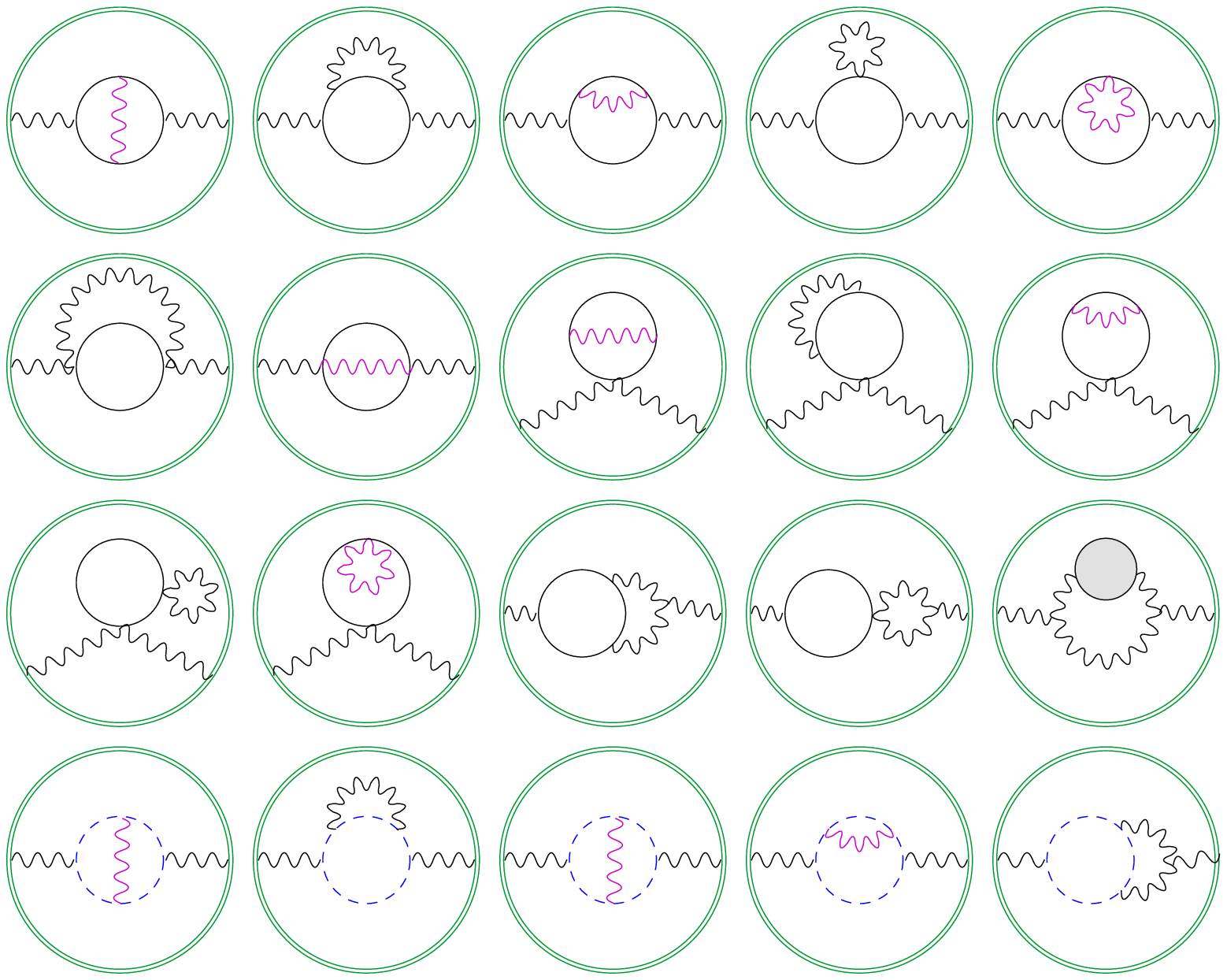}\\
\includegraphics[scale=0.45]{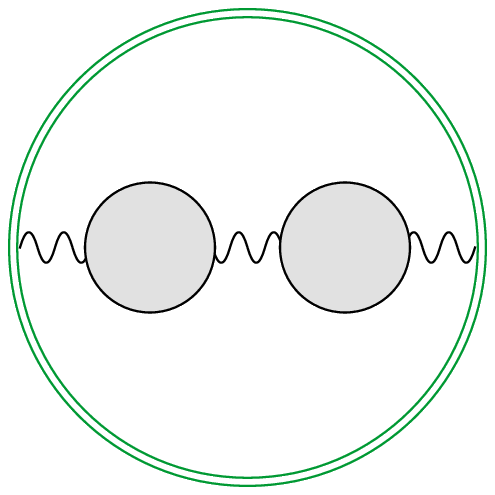}
\includegraphics[scale=0.6]{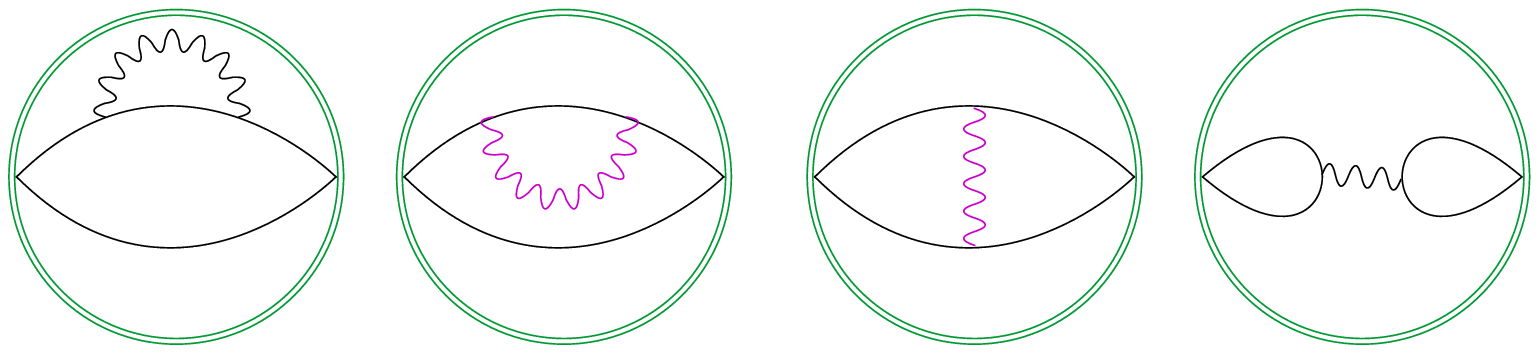}
\end{center}
\caption{\sl The Feynman diagrams contributing to order $\lambda^3$. They all have two vertices of
the Wilson loop along the contour $C$.}
\label{3-2}
\end{figure}
For the Feynman diagrams in fig. \ref{3-2}, one easily finds that each of them vanish separately.
For instance, for the second to the last diagram, the skeleton two-loop integral is given by
\begin{equation}
{1 \over N} \, 32\pi^2\, \mbox{Tr}M^2\cdot N\lambda^3 \, p^i\epsilon_{imn}\int_{k,l}\frac{k^m l^n}
{k^2l^2(k-p)^2(k-l)^2(l+p-k)^2}.
\end{equation}
Evidently, the two-loop integral should yield a result of the form:
\begin{equation}
A(p^2)p^m p^n +B(p^2)\eta^{mn}.
\end{equation}
Contracted with $p^i \varepsilon_{imn}$, this contribution vanishes identically.
Many of the diagrams in fig. \ref{3-2} vanish because self-energy of scalars and fermions are zero at one-loop.
\begin{figure}
\epsfxsize=2.2in
\begin{center}
\includegraphics[scale=0.7]{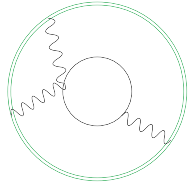}
\end{center}
\caption{\sl The Feynman diagram at order $\lambda^3$ that vanish by itself. It has three vertices of the
Wilson loop along the contour $C$.}
\label{3-3ggg-2}
\end{figure}
For the Feynman diagram in fig. \ref{3-3ggg-2}, the contribution is proportional to
\begin{equation}
\epsilon_{0mn}p_1^m\int_k\frac{k^n}{k^2(k-p_1)^2} = 0
\end{equation}
so vanishes identically.
\begin{figure}
\epsfxsize=2.2in
\begin{center}
\includegraphics[scale=0.7]{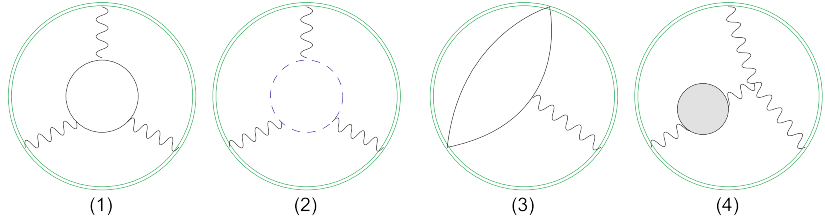}
\end{center}
\caption{\sl The Feynman diagrams at order $\lambda^3$ that cancel one another. They have three vertices of the Wilson loop along the contour $C$.}
\label{3-3cancel}
\end{figure}
\begin{figure}
\epsfxsize=2.2in
\begin{center}
\includegraphics[scale=0.7]{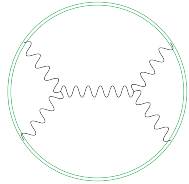}
\end{center}
\caption{\sl The Feynman diagram at order $\lambda^3$ that vanish identically. It has four maximal vertices of the Wilson loop along the contour.}
\label{3-4}
\end{figure}
The Feynman diagrams in fig. \ref{3-3cancel} cancel among themselves.
To see this, we need to manipulate loop integrals judiciously.
For instance, although
they contain a different number of epsilon tensors and gauge boson propagators, nontrivial cancelation occurs between the diagrams (2) and (3). The cancelation is possible because of various identities such as
\begin{equation}
(p_2^{m_1} p_1^{m_2}-\eta^{m_1m_2}p_1\cdot p_2)\epsilon_{0l_1m_1}p_1^{l_1}\epsilon_{0l_2m_2}
 p_2^{m_2} \,\, = \,\, -p_1^2p_2^2 + (\mbox{non-covariant terms}).
\end{equation}
In this way, two epsilon tensors cancel two gluon propagators in the diagram (2). It then cancels the diagram (3). The non-covariant terms vanish after the contour integration, as we have seen for two loop diagrams in subsection \ref{lambda2}. Through judicious manipulations, one can show all terms coming from the diagrams in fig. \ref{3-3cancel} cancel among themselves. We show details of the cancelation in Appendix \ref{calc}.

The Feynman diagram \ref{3-4} vanishes identically since it is proportional to $\epsilon^{00m}$.

\vspace{5mm}

Consider next the spacelike circle case.
Except the ones in fig. \ref{3-3cancel}, all other Feynman diagrams vanish by the same reason as for the
timelike line case, viz. either due to Tr$M=0$ or due to contraction of momenta with $\epsilon_{mnp}$ tensor.
After some manipulation, one also finds that all Feynman diagrams in fig. \ref{3-3cancel} vanish.
Start with the diagram (1). This gives a contribution proportional to
\begin{equation}
\int_{\tau_1>\tau_2>\tau_3}\dot{x}^{m_1}(\tau_1)\dot{x}^{m_2}(\tau_2)\dot{x}^{m_3}(\tau_3)
 \epsilon_{m_1l_1k_1}\epsilon_{m_2l_2k_2}\epsilon_{m_3l_3k_3}
 I^{l_1k_1l_2k_2l_3k_3}[x(\tau_1),x(\tau_2),x(\tau_3)],
\end{equation}
where $I^{l_1k_1l_2k_2l_3k_3}[x(\tau_1),x(\tau_2),x(\tau_3)]$
consists of integrals over positions of the interaction vertices.
One can always choose two epsilon tensors in the integrand and replace them with a sum of products of $\eta^{mn}$s. Then, after carrying out the position integration, the integral should contain terms
with one epsilon tensor whose indices are contracted with (a derivative of) $x^{m}(\tau)$. For instance,
it produces a term like
\begin{equation}
\dot{x}^{m_1}(\tau_1)\dot{x}^{m_2}(\tau_2)\dot{x}^{m_3}(\tau_3)\epsilon_{m_1m_2m_3} \, .
  \label{eg}
\end{equation}
For the spacelike circle, all $x^{m_i}(\tau_i)$s lie on $\mathbb{R}^2$. Therefore, all the terms like (\ref{eg}) vanish identically. By the same argument, the diagram (2) must vanish. Moreover,
this argument implies that any Feynman diagram with odd number of epsilon tensors ought to vanish once the contour integration is performed. Therefore, the diagrams (3) and (4) must vanish.

In summary, we find that three-loop contributions $W_3[C]$ vanish for both the timelike line case and the spacelike circle case.

\vspace{5mm}

\subsection{Diagrammatical proof of $W_{2n+1}[C]=0$}

\vspace{5mm}

Drawing from the lower order computations, we emphasize again that the cancelation observed among various Feynman diagrams is not specific to the ABJM theory. The same cancelation would persist even for a theory with a generic number of matter flavor multiplet, so long as the supersymmetry conditions Tr$M=$Tr$M^3=0$ hold. In fact, up to the order ${\cal O}(\lambda^3)$, only the gauge interaction vertices contributed to the Feynman diagrams. The quartet Yukawa interactions and the sextet scalar interactions specific to ${\cal N}=6$ supersymmetric ABJM theory will only start to contribute from the next order, viz. order ${\cal O}(\lambda^4)$. So, any result at order ${\cal O}(\lambda^4)$ or higher would be considered as the discriminating test of the ABJM theory against any others.

In this subsection, we prove one result on higher order terms: $W_{2n+1}[C]=0$ for {\sl all} $n$ as long as the contour $C$ lies inside $\mathbb{R}^2$. In other words, the Wilson loop expectation value receives nontrivial contribution only from even loop orders. Coincidentally, ABJM theory exhibits both infrared and ultraviolet divergences only at even loop orders. Here, however, we are considering not just these infinities but also finite
parts. We begin with the observation that any Feynman diagram can be drawn through the following two steps:

\vspace{3mm}

\hspace*{5mm}1. First, draw matter lines only. These diagrams need not be connected.

\hspace*{5mm}2. Next, add gluon lines.

\vspace{3mm}

In our proof, we shall follow these steps. First, we show that the diagrams with odd power of $\lambda$ vanish if there is no gluon propagator.
Then, by counting number of $\lambda$s and the $\varepsilon_{mnp}$ tensors as a single gluon propagator is added to a given diagram, we prove inductively that $W_{2n+1}[C]=0$.

\vspace{5mm}

Denote by $v_n$ the number of $n$-valent vertices in a given diagram $D$.
Here $v_1$ is the number of $A_m$ insertions from the Wilson loop, and $v_2$ is the number of $YY^\dag$ insertions from the Wilson loop. Evidently, the number $I$ of internal lines is
\begin{equation}
I = \frac12\sum_{n=1}^6 \, n \, v_n.
\end{equation}
The power $N_k$ of $\lambda$ is given as
\begin{eqnarray}
N_k &=& I-\sum_{n=3}^6v_n \nonumber \\
&=& \frac12(v_1+2v_2)+\frac12\sum_{n=3}^6(n-2)v_n.
\end{eqnarray}

\vspace{5mm}

\subsubsection{No gluon line}

\vspace{5mm}

If there is no gluon line in $D$, then the non-zero numbers are $v_2$, $v_4$ (Yukawa coupling) and $v_6$ (scalar potential).
In this case,
\begin{equation}
N_k = v_2+v_4+2v_6.
\end{equation}
Due to the supersymmetry conditions of the velocity matrix ${M_I}^J$, $v_2$ must be even in any non-vanishing diagram. Then we find that $v_4$ determines whether $N_k$ is even or odd.

Fermions must appear in $D$ as non-intersecting loops.
Let
\begin{equation}
v_4 = \sum n_i
\end{equation}
be a partition of $v_4$.
For each $n_i$, there is a fermion loop which consists of $n_i$ fermion propagators.
Since Yukawa terms do not contain gamma matrices, each loop has the trace of $n_i$ gamma matrices coming from the propagators.
If $n_i$ is odd, then the loop provides the epsilon tensor.
Therefore, if $v_4$ is odd, then there must be odd number of epsilon tensors.
We already proved that such diagram vanishes.
This shows that $W_{N_k=2n+1}[C]=0$ if there is no gluon line.

\vspace{5mm}

\subsubsection{Adding gluon lines}

\vspace{5mm}

We have shown that, if there is no gluon line, then $N_k$ and the number $N_\epsilon$ of epsilon tensors are equivalent modulo 2.
If we can show that the $N_k \equiv N_\epsilon$ mod 2 holds whenever a gluon line is added to a given diagram $D$ which already
satisfies that relation, then it proves $W_{2n+1}[C]=0$ in general by induction.
The above statement can be show to be true case by case.
For example, if a gluon line is added which connects two matter lines, then the number of vertices increases by two, the
number of propagators increases by three, which results in the increase of $N_k$ by one, and one epsilon tensor comes from
the gluon line.
Therefore, $N_k \equiv N_\epsilon$ mod 2 still holds.

There might be one subtle point in this argument.
Suppose that we have a generic diagram, and eliminate all the gluon lines.
Then the resulting diagram may contain a matter loop without any vertex, that is, there could be a matter loop to which only
gluons are attached.
Such a loop is not included in the discussion given in the previous subsection.
It seems that this causes no problem since the matter loop in question does not give no $k$ nor epsilon tensor.
This can be checked by considering a particular diagram with such a loop and eliminate the gluon lines.

\vspace{5mm}

Summarizing, we proved that $W_{2n+1}[C]=0$ to all orders in perturbation theory.
Note that the necessary ingredients for the proof are (1) Tr$M^{2n+1}=0$,
(2) $C$ lies in $\mathbb{R}^2 \in \mathbb{R}^4$, and (3) the action is classically conformally invariant
so that the interaction terms
consist only of gauge, quartic Yukawa and sextet scalar couplings. It should be noted that our proof
is insensitive to specific value of the coefficients for each terms in the Yukawa couplings
and the scalar potential.

\section{Reduction to the Gaussian Matrix Model} \label{Matrix Model}
In this section, we revisit planar perturbative evaluation of the circular Wilson loop.
As demonstrated in the previous section, expectation value of the circular Wilson loop
${\cal W}_{\bf N} [C, M]$ contains features similar to the circular Wilson loop of 4-dimensional
${\cal N}=4$ super Yang-Mills theory \cite{Erickson:2000af}.
The first feature is that, apart from quantum corrections to gauge and scalar propagators,
all other diagrams vanish. Second feature is that sum of quantum corrected gauge and scalar propagators
is reduced to a constant up to total derivative term.
It is remarkable that these features persists despite field contents,
interactions and even spacetime dimensions are different between the two conformal field theories.
A new feature that arose for the ABJM theory was that there is an extra contribution
from Chern-Simons interactions. Combining all these features, we expect that expectation value
of a circular ABJM Wilson loop computed in planar perturbation theory can be brought into the form
\begin{align}
 \langle {\cal W}_{\bf N} [C, M] \rangle= \langle {\cal W}_{\bf N} [C] \rangle_{\rm CS}
\langle {\cal W}_{\bf N} [C, M] \rangle_{\rm ladder} \, .
\label{ansatz}
\end{align}
Here, $\langle {\cal W}_{\bf N} [C] \rangle_{\rm CS}$ denotes expectation value of an unknotted Wilson loop
in {\sl pure} Chern-Simons theory, while $\langle {\cal W}_{\bf N} [C, M] \rangle_{\rm ladder}$
is the ladder part \footnote{This means ladders of quantum corrected propagators.} of the Wilson loop
in the ABJM theory.

In this section, we employ the factorized form \eqref{ansatz} as an ansatz, and explore possible exact results
concerning the Wilson loop.

\subsection{Gaussian Matrix Model}
In extracting Gaussian matrix model, our starting point will be the following assumptions.
\begin{itemize}
 \item The two point function of gauge boson$+$scalar bilinear is a constant.
 \item The Wilson loop has the structure \eqref{ansatz}. All other diagrams than those in \eqref{ansatz} cancel each other and do not contribute to the expectation value.
\end{itemize}
These assumptions are actually true up to the order of ${\cal O}(\lambda^3)$,
as we have demonstrated in the previous section.

Let us first consider the ladder part.  In the planar limit, by large-$N$ factorization,
expectation value  $\langle {\cal W}_{\bf N} [C, M] \rangle$ is given by
\bea
\langle {\cal W}_{\bf N}[C, M] \rangle = \langle W_{\bf N} [C, M] \rangle
= \langle \overline{W}_{\overline {\bf N}} [C, M] \rangle.
\eea
Parallel to ${\cal N}=4$ super Yang-Mills theory \cite{Erickson:2000af},
we now show that $\langle {\cal W}_{\bf N}[C, M]\rangle_{\rm ladder}$ can be related
to the Gaussian matrix model, but with an interesting twist.

Consider the Wilson loop on $\mathbb{R}^3$ and recall the exponential $\Phi(\tau)$
\begin{align}
 \Phi(\tau):= [i A_{m}(x)\dot{x}^{m}(\tau)+M^{I}{}_{J}(Y^{I}Y^{\dag}_{J})(x)]_{x = x(\tau)}.
\end{align}
This $\Phi(\tau)$ is in adjoint of $U(N)$, so we expanded it in the basis of Lie algebra generators $T^a$:
\begin{align}
 \Phi(\tau)=\Phi^{a}(\tau)T_{a}.
\end{align}
From the assumption, the two point function of $\Phi$ is a constant. This is supported by the
two loop result \eqref{importantresult}:
\begin{align}
 \langle\Phi^{a}(\tau_1)\Phi^{b}(\tau_2)\rangle=\frac{\delta^{ab}}{N} f(\lambda)
=\frac{\delta^{ab}}{N}(\lambda^2 +O(\lambda^4)) \label{flambda}.
\end{align}
Therefore, contribution to the Wilson loop expectation value from ladder diagrams of quantum
corrected $\Phi^a$ propagators is obtainable from
\begin{align}
 \langle {\cal W}_{\bf N}[C, M]\rangle_{\rm ladder}
={1 \over N} \sum_{n=0}^{\infty}\int_{\tau_1>\tau_2>\dots>\tau_n}
\Tr\left[T_{a_1}\dots T_{a_n}\right]\langle \Phi^{a_1}(\tau_1)\dots \Phi^{a_n}(\tau_n)\rangle_{\rm ladder} .
\end{align}
By the assumption, the integrands in the above equation are $\tau$ independent,
so $\tau$ integral just yield angular volume as
\begin{align}
 \langle {\cal W}_{\bf N}[C, M]\rangle_{\rm ladder}
={1 \over N} \sum_{n=0}^{\infty}\frac{1}{n!}(2\pi)^n
\Tr\left[T_{a_1}\dots T_{a_n}\right]\langle \Phi^{a_1}\dots \Phi^{a_n}\rangle_{\rm ladder}.\label{Expand Phi}
\end{align}
Here, the expectation values $\langle \cdots \rangle_{\rm ladder}$ are evaluated according to the Wick's theorem using the propagator \eqref{flambda}. We now rewrite the above series in a simpler form. Introduce $N^2$ real variables $X^a$ and the Gaussian integral $\langle F(X) \rangle_{\rm mm}$ for a function $F(X)$ as
\begin{align}
 \langle F(X)\rangle_{\rm mm}:=&\frac{1}{Z}\int d^{N^2}X
\, F(X) \exp\left[-\frac12\frac{N}{(2\pi)^2 f(\lambda)}\sum_{a}X^{a}X^{a} \right],\label{Gaussian integral}\\
Z:=&\int d^{N^2}X \exp\left[-\frac12\frac{N}{(2\pi)^2 f(\lambda)}\sum_{a}X^{a}X^{a} \right].
\end{align}
The Wick contracted expectation values can be replaced by the Gaussian integral.
This brings \eqref{Expand Phi} to the form
\begin{align}
 \langle {\cal W}_{\bf N}[C, M]\rangle_{\rm ladder}=&{1 \over N} \sum_{n=0}^{\infty}\frac{1}{n!}
\Tr\left[T_{a_1}\dots T_{a_n}\right]\langle X^{a_1}\dots X^{a_n}\rangle_{\rm mm}\nonumber\\
=&\left\langle{1 \over N} \Tr\left(e^{X}\right)\right\rangle_{\rm mm},\label{matrix model expectation value}
\end{align}
where we introduced a single Hermitian matrix $X:=X^{a}T_{a}$. The Gaussian integral \eqref{Gaussian integral} can then be rewritten as a Gaussian matrix integral:
\begin{align}
\langle F(X)\rangle_{\rm mm}=&\frac{1}{Z}\int d^{N^2}X
\, F(X) \exp\left[-\frac{N}{(2\pi)^2 f(\lambda)}\Tr(X^2) \right].\label{Gaussian matrix model}
\end{align}
In the planar limit, the expectation value \eqref{matrix model expectation value} can be evaluated
in terms of modified Bessel function $I_1$ as
\begin{align}
 \langle {\cal W}_{\bf N}[C, M]\rangle_{\rm ladder}
 =\left\langle  {1 \over N} \Tr\left(e^{X}\right)\right\rangle_{\rm mm}
=\frac{1}{\pi\sqrt{2f(\lambda)}}I_{1}(2\sqrt{2}\pi \sqrt{f(\lambda)}).
   \label{interpolate}
\end{align}
In the large $f(\lambda)$ limit, we obtain asymptote of the Wilson loop expectation value as
\begin{align}
\langle {\cal W}_{\bf N}[C, M]\rangle_{\rm ladder}\sim \exp(2\sqrt{2}\pi\sqrt{f(\lambda)})
\label{large}
\end{align}
up to computable pre-exponential factors. If large $f(\lambda)$ limit is also large $\lambda$ limit,
this is a prediction of the ABJM theory that could be compared with the string theory dual.

\subsection{Chern-Simons Contribution}
In ABJM theory, the Wilson loop expectation value \eqref{ansatz} contains an additional contribution
from {\sl pure} Chern-Simons interactions. We need to examine large $f(\lambda)$ limit of this contribution
as well. By assumption we take $\langle W_{\bf N} [C] \rangle_{\rm CS}$ is the same as unknotted Wilson loop expectation value in pure Chern-Simons theory. Exact answer of the latter is known \cite{Witten:1988hf}:
\begin{align}
 \langle {\cal W}_{\bf N} [C] \rangle_{\rm CS}
 ={1 \over N} \frac{q^{\frac{N}{2}}-q^{-\frac{N}{2}}}{q^{\frac{1}{2}}-q^{-\frac{1}{2}}},
\qquad q:=\exp\left(\frac{2\pi i}{k+N}\right).
\end{align}
In the 't Hooft limit, this becomes
\begin{align}
 \langle {\cal W}_{\bf N} [C] \rangle_{\rm CS}= \frac{1+\lambda}{\pi \lambda}
\sin \frac{\pi \lambda}{1+\lambda}.
\end{align}
We see that large $\lambda$ asymptote is given by
\begin{align}
  \langle {\cal W}_{\bf N} [C] \rangle_{\rm CS}=  (\lambda^{-1}+\dots).
\end{align}
We see that this contribution yields exponentially small corrections compared to the ladder diagram contribution \eqref{large}. The $\lambda^{-1}$ asymptote still carries an interesting information, since it changes leading power of $\lambda$ in the pre-exponential. In particular, this indicates that number of zero-modes of the string worldsheet configuration dual to the circular Wilson loop in the ABJM theory is different from that in the ${\cal N}=4$ super Yang-Mills theory.

\subsection{Interpolation between Weak and Strong Coupling}
By AdS/CFT correspondence, large $\lambda$ behavior of ${\cal W}_{\bf N} [C, M]$ was determined from minimal surface configuration of the string worldsheet in section \ref{IIA}.
On the other hand, small $\lambda$ behavior of ${\cal W}_{\bf N} [C, M]$ was determined from planar perturbation theory in section \ref{perturbative}. This poses an interesting question: what kind of function $f(\lambda)$ can interpolate between the weak
and strong coupling behavior?
We assume that (\ref{interpolate}) can be used for this purpose with a suitable choice of $f(\lambda)$.
The small $\lambda$ behavior of $f(\lambda)$ can be obtained by comparing (\ref{flambda}) with (\ref{importantresult}), and the result is already given in (\ref{flambda}).
{\sl Assuming} that large $\lambda$ limit is also large $f(\lambda)$ limit, the large $\lambda$ behavior can be extracted by comparing (\ref{large}) with (\ref{AdS Wilson loop result 1}). We obtain
\begin{equation}
f(\lambda) \to \left\{
\begin{array}{cc}
{\lambda^2} & (\lambda\to0) \\ [2mm]
\displaystyle{\frac\lambda{4}} & (\lambda\to\infty)
\end{array}
\right.  \, .
\end{equation}
When comparing various physical observables at weak coupling limit from the ABJM theory and at strong coupling limit from the AdS$_4 \times \mathbb{CP}^3$ string theory, various interpolating functions analogous to $f(\lambda)$ were introduced. An interesting question is whether some of these interpolating functions are actually the same one. To test this possibility, consider the interpolating function $f(\lambda)$ introduced in the context of the giant magnon spectra \cite{Gaiotto:2008cg,Nishioka:2008gz,Grignani:2008is}. There, it was noted that dispersion relation of AdS$_4$ giant magnon takes exactly the same form as that of AdS$_5$ giant magnon except that ${\cal N}=4$ super Yang-Mills `t Hooft coupling $g^2 N$ is now replaced by a nontrivial interpolating function $h(\lambda)$ of the ${\cal N}=6$ superconformal Chern-Simons `t Hooft coupling:
\bea
g^2 N \Big\vert_{\rm SYM} \rightarrow 16 \pi h^2(\lambda) \Big\vert_{\rm ABJM}.
\eea
At weak coupling, $h(\lambda) \sim \lambda$. So, it is encouraging that the interpolating function associated with the giant magnon and the interpolating function associated with the circular Wilson loop are relatable
each other as $h^2(\lambda)=f(\lambda)$. But it seems this would not work for all coupling regime because $h(\lambda)$ actually interpolates as
\begin{equation}
h(\lambda) \to \left\{
\begin{array}{cc}
\lambda & (\lambda\to0) \\ [2mm]
\displaystyle{\sqrt{\frac\lambda{2}}} & (\lambda\to\infty) \, .
\end{array}
\right.
\end{equation}
We see that it behaves differently at the strong coupling regime,
so the two interpolating functions are not identifiable.
Our proposal of the Gaussian matrix model suggests that there ought to be an independent
interpolating function $f(\lambda)$ specific to the circular Wilson loop observable.
Since $f(\lambda)$ summarizes all-order corrections to the vacuum polarization of the ABJM gauge fields,
interpolating functions that would enter static quark potential or total cross section of 2-body boson
or fermion matter might be related to $f(\lambda)$. It would be very interesting to clarify the relation,
if any, and compute higher order terms of $f(\lambda)$.

\section{Discussions} \label{discuss}
In this section, we discuss several interesting issues left for future investigation.

We identified an elementary Wilson loop ${\cal W}_{\bf N} [C, M]$
which transforms correctly under generalized time-reversal, and we proposed
that this is dual to fundamental Type IIA string.
Though the identification is correct from the viewpoint of charge conservation
and time-reversal symmetry, consideration of other symmetries remains to be understood better.
For the Wilson loop, there exists a unique supersymmetric configuration and it preserves
${1 \over 6}$ of the ${\cal N}=6$ superconformal symmetry.
On the other hand, the fundamental string in AdS$_4$ preserves ${1 \over 2}$ supersymmetry.
Related, the supersymmetric Wilson loop preserves SU(2)$\times$SU(2) subgroup of the SU(4) R-symmetry,
while the supersymmetric fundamental string in AdS$_4$ preserves SU(3) subgroup.
We also observed that string configuration preserving ${1 \over 6}$ supersymmetry
and SU(2) subgroup is obtainable by smearing string position in $\mathbb{CP}^3$ over a $\mathbb{CP}^1$.
Still, given that a fundamental string preserving ${1 \over 2}$ supersymmetry and SU(3) subgroup of SU(4)
R-symmetry exists, a supersymmetric Wilson loop with the same symmetry is yet to be identified.

With the Wilson loop and its holographic dual is identified, various physical observables are computable.
By inspection, static quark potential at conformal point is exactly the same as AdS$_5$
and ${\cal N}=4$ super Yang-Mills counterpart. It would be interesting to extend the computation
to Coulomb branch and compared the two sides. Also, various lightlike Wilson loops and their cusp
anomalous dimensions can be computed.  It would be interesting to see if they are related to
scattering amplitudes and the fermionic T-duality of the ABJM theory.

Another important direction is to compute the ${\cal O}(\lambda^4)$ contribution
to the circular Wilson loop. The computation will elucidate validity of the factorization hypothesis
of the Wilson loop expectation value in terms of Gaussian matrix model proposed in section 7.
The computation is also a nontrivial test of ${\cal N}=6$ supersymmetry since, from this order,
Feynman diagrams involving Yukawa coupling and sextet scalar interactions specific to the ABJM theory
begin to contribute. So we could find some distinguished features of ${\cal N}=6$ ABJM model from
${\cal N}=2$ superconformal Chern-Simons models. At the same time checking the cancellation is
highly non-trivial interesting problem.
In ${\cal N}=4$ super Yang-Mills case, the reduction from circular Wilson loop to the Gaussian
matrix model is proved using localization\cite{Pestun:2007rz}. Similar derivation for the circular
Wilson loop in the ABJM theory is also an interesting problem.

We intend to report progress of these issues in forthcoming publications.

\section*{Acknowlegement}
We are grateful to Fernando Alday, Dongsu Bak, Lance Dixon, Andreas Gustavsson and Juan Maldacena for enlightening discussions on issues related to this work. We also acknowledge pertinent conversations with Dongmin Gang, Eunkyung Koh and Jaesung Park.
This work was supported in part by SRC-CQUeST-R11-2005-021, KRF-2005-084-C00003, EU FP6 Marie Curie Research \& Training Networks MRTN-CT-2004-512194 and HPRN-CT-2006-035863 through MOST/KICOS and F.W. Bessel Award of Alexander von Humboldt Foundation (SJR).

\appendix

\section{Notation, Convention and Feynman Rules \label{notation}}
\subsection{Notation and Convention}
$\bullet$ $\mathbb{R}^{1,2}$ metric:
\bea
&& g_{mn} = \mbox{diag}(-, +, +) \quad \mbox{with} \quad  m,n = 0, 1, 2. \nonumber \\
&& \epsilon^{012} = - \epsilon_{012} = +1 \nonumber \\
&& \epsilon^{mpq} \epsilon_{mrs} = - (\delta^p_r \delta^q_s - \delta^p_s \delta^q_r); \qquad
\epsilon^{mpq} \epsilon_{mpr} = - 2 \delta^q_r \nonumber \\
\eea
$\bullet$ $\mathbb{R}^{1,2}$ Majorana spinor and Dirac matrices:
\bea
&& \psi \equiv \mbox{two-component} \,\,\, \mbox{Majorana} \,\,\, \mbox{spinor} \nonumber \\
&& \psi^\alpha = \epsilon^{\alpha \beta}\psi_\beta ,
\quad \psi_\alpha = \epsilon_{\alpha \beta} \psi^\beta \quad
\mbox{where} \quad \epsilon^{\alpha \beta} = - \epsilon_{\alpha \beta} = i \sigma^2 \nonumber \\
&& {\gamma^m_\alpha}^\beta = ( i \sigma^2, \sigma^3, \sigma^1),
\quad (\gamma^m)_{\alpha \beta} = (-\mathbb{I}, \sigma^1, -
\sigma^3) \quad \mbox{obeying} \quad \gamma^m \gamma^n = g^{mn} -
\epsilon^{mnp} \gamma_p. \eea

\subsection{ABJM Theory}
$\bullet$ Gauge and global symmetries:
\bea
&& \mbox{gauge symmetry}: \quad \mbox{U(N)} \cdot \overline{\mbox{U(N)}} \nonumber \\
&& \mbox{global symmetry}: \quad \mbox{SU(4)}
\eea
We denote trace over U(N) and $\overline{\rm U(N)}$ as Tr and $\overline{\rm Tr}$, respectively.
We also denote generators for U($N$) and $\overline{\rm U}(N)$ gauge groups by the same notation $T^a, (a=0, 1, \cdots, N^2-1)$. They are Hermitian and normalized to
\bea
\mbox{Tr} (T^a T^b) = {1 \over 2} \delta^{ab}.
\eea

$\bullet$ On-shell fields are gauge fields, complexified Hermitian scalars and Majorana spinors ($I=1,2,3,4$):
\bea
&& A_m: \quad \mbox{Adj}\,\,\, (\mbox{U(N)}); \hskip2cm \overline{A}_m : \quad
\mbox{Adj}\,\,\, \overline{\mbox{U(N)}} \nonumber \\
&& Y^I = (X^1 + i X^5, X^2 + i X^6, X^3 - i X^7, X^4 - i X^8):  \qquad ({\bf N}, \overline{\bf N}; {\bf 4}) \nonumber \\
&& Y^\dagger_I = (X^1 - i X^5, X^2 - i X^6, X^3 + i X^7, X^4 + i X^8): \hskip0.8cm ( \overline{\bf N}, {\bf N}; \overline{\bf 4})
\nonumber \\
&& \Psi_I = (\psi^2 + i \chi^2, - \psi^1 - i \chi^1, \psi_4  - i \chi_4 , - \psi_3 + i \chi_3 ) : \hskip0.4cm
({\bf N}, \overline{\bf N}; \overline{\bf 4}) \nonumber \\
&& \Psi^{\dagger I} = (\psi_2 - i \chi_2, - \psi_1 + i \chi^1, \psi^4 + i \chi^4, - \psi^3 - i \chi^3): \hskip0.3cm
(\overline{\bf N}, {\bf N}; {\bf 4})
\eea
$\bullet$ action: To suppress the cluttering $2\pi$ factors, we use the notation $\kappa:= {k\over 2\pi}$.
\bea
I &=& \kappa   \int_{\mathbb{R}^{1,2}}
\Big[ \, \epsilon^{mnp} \mbox{Tr} \left(\frac12 A_m \partial_n A_p +{ i \over 3} A_m A_n A_p \right)
-  \epsilon^{mnp} \overline{\mbox{Tr}} \left( \frac12 \overline{A}_m \partial_n \overline{A}_p +{ i \over 3} \overline{A}_m
\overline{A}_n \overline{A}_p \right) \nonumber \\
&& \hskip1cm +{1 \over 2} \overline{\mbox{Tr}} \left( -(D_m Y)^\dagger_I D^m Y^I  + i \Psi^{\dagger I} D \hskip-0.22cm / \Psi_I  \right) + {1 \over 2} \mbox{Tr} \left(- D_m Y^I (D^m Y)^\dagger_I  +
 i \Psi_I D \hskip-0.22cm / \Psi^{\dagger I}  \right)
\nonumber \\
&&
\hskip1cm - V_{\rm F} - V_{\rm B} \, \Big]
\eea
Here, covariant derivatives
are defined as
\bea
D_m Y^I = \partial_m Y^I + i A_m Y^I - i Y^I \overline{A}_m \, , \quad D_m Y^\dagger_I = \partial_m Y^\dagger_I + i \overline{A}_m Y^\dagger_I - i Y^\dagger_I A_m
\eea
and similarly for fermions $\Psi_I, \Psi^{\dagger I}$. Potential terms are
\bea
V_{\rm F} &=& i\, \overline{\mbox{Tr}} \Big[ Y^\dagger_I Y^I \Psi^{\dagger J} \Psi_J
- 2 Y^\dagger_I Y^J  \Psi^{\dagger I} \Psi_J  + \epsilon^{IJKL} Y^\dagger_I \Psi_J Y^\dagger_K \Psi_L]
\nonumber \\
&-& i\, \mbox{Tr} [Y^I Y^\dagger_I \Psi_J  \Psi^{\dagger J}
- 2 Y^I Y^\dagger_J \Psi_I \Psi^{\dagger J} + \epsilon_{IJKL} Y^I \Psi^{\dagger J} Y^K \Psi^{\dagger L} \Big]
\eea
and
\bea
V_{\rm B} &=& - {1 \over 3}
\overline{\mbox{Tr}} \Big[ \, Y^\dagger_I Y^J Y^\dagger_J Y^K Y^\dagger_K Y^I
+ Y^\dagger_I Y^I Y^\dagger_J Y^J Y^\dagger_K Y^K \nonumber \\
&& \hskip1.8cm + 4 Y^\dagger_I Y^J Y^\dagger_K
Y^I Y^\dagger_J Y^K -
6 Y^\dagger_I Y^I Y^\dagger_J Y^K Y^\dagger_K Y^J \, \Big]
\eea
At quantum level, since the Chern-Simons term shifts by integer multiple of
$8 \pi^2$, not only $N$ but also $k$
should be integrally quantized.
At large $N$, we expand the theory and
physical observables in double series of
\bea
g_{\rm st} = {1 \over N}, \qquad \lambda = {N \over k} =
 {N \over 2 \pi\kappa}
\eea
by treating them as continuous perturbation parameters.

\subsection{Feynman Rules}
$\bullet$ We adopt Lorentzian Feynman
rules and manipulate all Dirac matrices and $\epsilon_{mnp}$
tensor expressions to scalar integrals.
For actual evaluation of these integrals,
we shall go the Euclidean space integral by the Wick rotation,
which corresponds to $x^0 \rightarrow -i\tau$. In the momentum
space, this means we change the contour of $p_0$
 to the imaginary axis 
following
the standard Wick rotation. Then in terms
of integration measure, we simply replace
${\rm d}^{2 \omega} k \rightarrow
 i {\rm d}^{2 \omega} k_{\rm E}$ together
 with $p^2 \rightarrow  + p_{\rm E}^2$.
The procedure is known to obey Slavnov-Taylor
identity, at least to two loop order.
\hfill\break
$\bullet$ We choose covariant gauge fixing condition for both gauge groups:
\bea
\partial^m A_m = 0 \qquad \mbox{and} \qquad \partial^m \overline{A}_m = 0
\eea
and work in Landau gauge.
Accordingly , we introduce a pair of
Faddeev-Popov ghosts $c, \overline{c}$ and their conjugates, and add to $I$ the ghosts action:
\bea
I_{\rm ghost} = \kappa \int_{\mathbb{R}^{1,2}} \Big[ \mbox{Tr} \partial^m c^* D_m c + \overline{\mbox{Tr}}
\partial^m  \overline{c}^* D_m \overline{c} \Big]
\eea
Here, $D_m c = \partial_m c + i [A_m, c]$ and $D_m \overline{c} = \partial_m \overline{c} + i
[\overline{A}_m, \overline{c}]$.

$\bullet$ Propagators in U(N)$\times\overline{\mbox{U(N)}}$ matrix notation:
\bea
\mbox{gauge propagator}: \quad && \Delta_{mn}(p) =\kappa^{-1} \mathbb{I} \,  {\epsilon_{mnr} p^r \over p^2 - i \epsilon} \nonumber \\
\mbox{scalar propagator}: \quad && {D_I}^J (p) = \kappa^{-1} \delta_I^J \,  {-i \over p^2 - i \epsilon}  \nonumber \\
\mbox{fermion propagator}: \quad && {S^I}_J (p) =\kappa^{-1} \delta^I_J \, { i p \hskip-0.22cm / \over p^2 - i \epsilon}   \nonumber \\
\mbox{ghost propagator}: \quad && K(p) \, = \, \kappa^{-1} {-i \over p^2 - i \epsilon}
\eea
$\bullet$ Interaction vertices are obtained by multiplying $i = \sqrt{-1}$ to nonlinear terms of the Lagrangian density. Note that the paramagnetic coupling of gauge fields to scalar fields has the invariance property under simultaneous exchange between $A_m, Y^I$ and $\overline{A}_m, Y^\dagger_I$.

$\bullet$ Momentum representation:

\begin{equation}
\int_p := \int\frac{d^3p}{(2\pi)^3},
\end{equation}
\begin{equation}
Y^I(x) = \int_pe^{ip\cdot x}Y^I(p)
\end{equation}
\begin{equation}
YY^\dag(p) := \int_{q}M_I{}^JY^I(q+p)Y^\dag_J(q).
\end{equation}

\vspace{1cm}

\section{Supersymmetry condition for generic contour \label{proof}}

\vspace{5mm}

Consider the generalized supersymmetry conditions for the Wilson loop:
\begin{equation}
\xi_{IJ}n\hspace{-2mm}/(\tau)+M_I{}^K(\tau)i\xi_{KJ} = 0.
\end{equation}
We assume that $n(\tau)^2=-1$ and $M_I{}^J(\tau)$ is a hermitian matrix.
$M(\tau)$ can be decomposed as
\begin{equation}
M(\tau) = U^\dag(\tau)\Lambda(\tau)U(\tau),
\end{equation}
where $\Lambda(\tau)=\Lambda$ is a constant diagonal matrix unless $U(\tau)$ is allowed to have discontinuities.
We assume that $U(\tau)$ is continuous, and
$\Lambda=\mbox{diag}(-1,-1,+1,+1)$ so that the generalized supersymmetry conditions have a non-trivial solution.

Let us assume, without loss of generality, that $U(\tau=0)$ is the identity matrix.
Then, non-zero components of $\xi_{IJ}$ are $\xi_{12}$ and $\xi_{34}$ with
\begin{equation}
\xi_{12}n\hspace{-2mm}/ = i\xi_{12}, \hspace{5mm} \xi_{34}n\hspace{-2mm}/ = -i\xi_{34},
   \label{ev}
\end{equation}
where $n^m=n^m(0)$.
Define
\begin{equation}
\xi_{IJ}(\tau) := U(\tau)_I{}^K\xi_{KJ}.
\end{equation}
Notice that $\xi_{IJ}(\tau)$ is no longer anti-symmetric.
This satisfies
\begin{equation}
\xi_{IJ}(\tau)n\hspace{-2mm}/(\tau)+\Lambda_I{}^K(\tau)i\xi_{KJ}(\tau) = 0.
\end{equation}
Since $n^m(\tau)$ is related to $n^m$ by a Lorentz transformation $L^m{}_l(\tau)$, we find its spinor representation
\begin{equation}
S(\tau)\gamma_m S^{-1}(\tau) = \gamma_m L^m{}_l(\tau)n^l,
\end{equation}
with $S(0)=1$ by definition.
Now the supersymmetry condition becomes
\begin{equation}
\xi_{IJ}(\tau)S(\tau)n\hspace{-2mm}/+\Lambda_I{}^Ki\xi_{KJ}(\tau)S(\tau) = 0.
\end{equation}

Consider the following condition:
\begin{equation}
\xi_{13}(\tau)S(\tau)n\hspace{-2mm}/-i\xi_{13}(\tau)S(\tau) = 0
\end{equation}
for a generic $\tau$.
Since $\xi_{13}(\tau)=U_1{}^4(\tau)\xi_{43}$, this condition implies
\begin{equation}
U_1{}^4(\tau)=0 \hspace{5mm} \mbox{or} \hspace{5mm}
\xi_{34}S(\tau)n\hspace{-2mm}/-i\xi_{34}S(\tau) = 0.
\end{equation}
As long as the contour is smooth, that is, $S(\tau)$ is continuous,
the latter contradicts with (\ref{ev}).
Therefore, we conclude that $U_1{}^4(\tau)=0$ holds.
Using similar arguments, one can show that
\begin{equation}
U(\tau) = \left[
\begin{array}{cc}
u_1(\tau) & 0 \\ 0 & u_2(\tau)
\end{array}
\right],
\end{equation}
where $u_1(\tau),u_2(\tau)$ are $U(2)$ matrices.
This immediately implies that $M(\tau)=\Lambda$ and $n^m(\tau)=n^m$.

\vspace{1cm}

\section{Vacuum polarization \label{polarization}}

The self-energy correction enters in the same form for
the U($N$) and the $\overline{{\rm U}(N)}$ gauge fields.
Therefore we focus on the correction to $A$ gauge
field only. At the one-loop level,
the boson, the fermion, the gauge and the ghost loops
may in general contribute to the gauge self-energy correction.
In this appendix, we identify these self-energy contributions.

We begin with the scalar loop contribution. It is the sub-diagram of
Fig.~\ref{pse1}. The momentum $k$ plays the role of the external momentum.
The self energy contribution reads
\bea
 i\, \Pi^s_{ab} (k)= (i)^2 [i\,]^2 (4)i
 \int {{\rm d}^{2 \omega} \ell \over (2 \pi)^{2 \omega}}
{ (2\ell +k)_a (2\ell+k)_b
  \over (k+\ell)^2\,\,  \ell^2}\,,
\eea
where the extra factor $4$ comes from the fact that
4 complex scalars are coupled to the gauge field.
Using the dimensional regularization, one obtains
\be
i\,\Pi^s_{ab} (k)= (4)i\Bigl[  {{ k_a  k_b -g_{ab} k^2 }\over 16 k}\Bigr]\,.
\ee

Similarly, for the fermion loop, the self-energy
contribution becomes
\bea
 i\, \Pi^f_{ab} (k)= (i)^2 [i\,]^2 (4)(-)_{\rm FD}
 \, i \int {{\rm d}^{2 \omega} \ell \over (2 \pi)^{2 \omega}}
{ \tr \gamma_a \, (\ell\hskip-0.23cm / +
k\hskip-0.23cm / \, ) \,\gamma_b\,  \ell \hskip-0.23cm /
  \over (k+\ell)^2\,\,  \ell^2}\,,
\eea
where again the extra factor four comes from the fact that there are
4 complex fundamental fermions. Using the $\gamma$ matrix identity and the
 dimensional regularization, the contribution becomes
 \be
i\, \Pi^f_{ab} (k)= (4) i\Bigl[  {{ k_a  k_b -g_{ab} k^2 }\over 16k}\Bigr]\,.
\ee
Hence, each complex matter contributes by the same weight and sign.

One can continue the dimensions $2\omega$ to four and obtain the vacuum polarization
in four-dimensional Yang-Mills theories. The integration leads to the logarithmic
divergence in this case contributing positively to the $\beta$-function of the Yang-Mills
coupling. Again, boson and fermion contributions add up.

For the gluon self-energy contribution, we have
 \bea
i\, \Pi^A_{ab} (k) &=& (3)\cdot (3) [i^2]
\Bigl[{i\kappa \over 3}\Bigr]^2 \Bigl[{1\over \kappa}\Bigr]^2
(i)^2 [i\,]^2 (4) i
 \int {{\rm d}^{2 \omega} \ell \over (2 \pi)^{2 \omega}}
{ \epsilon^{mbn} \epsilon^{jai} \epsilon_{imq} \epsilon_{njr} (\ell+k)^q \ell^r
  \over (k+\ell)^2\,\,  \ell^2}\,,\nonumber\\
&=&
 i \int {{\rm d}^{2 \omega} \ell \over (2 \pi)^{2 \omega}}
{(\ell +k)_a \ell_b + (\ell +k)_b \ell_a
  \over (k+\ell)^2\,\,  \ell^2}\,.
\eea
It becomes
\be
i\, \Pi^A_{ab} (k)=  -i\Bigl[  {{ k_a  k_b +g_{ab} k^2 }\over 32 k}\Bigr]\,,
\ee
which alone does not respect the gauge invariance.
However, there exists also the ghost loop contribution,
\bea
i\, \Pi^{\rm gh}_{ab} (k)
=(i)^2 [i\,]^2 (-)i \int {{\rm d}^{2 \omega} \ell \over (2 \pi)^{2 \omega}}
{(\ell +k)_a \ell_b + (\ell +k)_b \ell_a
  \over (k+\ell)^2\,\,  \ell^2}\,,
\eea
where we put the extra ($-$) sign due to the ghost statistics.
Therefore, the ghost contribution cancels out precisely
the gauge loop contribution, reproducing the well-established result~\cite{Chen:1992ee}.

Again, analytically continuing to four dimensions, the integral expression for the
gauge part changes while the ghost integral remains intact.
With Yang-Mills couplings, both contributions no longer cancel each other but
contribute negatively to the $\beta$-function.

\vspace{1cm}

\section{Feynman integrals for 3-loop diagrams in figure \ref{3-3cancel} \label{calc}}
\vspace{5mm}

The diagram (2) provides
\begin{equation}
 -32\pi^3\lambda^3E_{m_1m_2m_3}\int_k\frac1{k^2(k-p_1)^2(k-p_1-p_2)^2}\left[ T^{m_1m_2m_3}_{F1}
   +T^{m_1m_2m_3}_{F2}+T^{m_1m_2m_3}_{F3}+T^{m_1m_2m_3}_{F4} \right]
\end{equation}
where we defined
\begin{equation}
E_{m_1m_2m_3} = \prod_{i=1}^3\frac{\epsilon_{0l_im_i}p_i^{l_i}}{p_i^2},
\end{equation}
and
\begin{eqnarray}
T^{m_1m_2m_3}_{F1} &=&-8k^{m_1}(k-p_1)^{m_2}k^{m_3}, \\
T^{m_1m_2m_3}_{F2} &=&-2(p_2^{m_1}p_1^{m_2}-\eta^{m_1m_2}p_1\cdot p_2)k^{m_3}
                            -2(p_3^{m_2}p_2^{m_3}-\eta^{m_2m_3}p_2\cdot p_3)k^{m_1} \nonumber \\
                         & &-2(p_1^{m_3}p_3^{m_1}-\eta^{m_3m_1}p_3\cdot p_1)(k-p_1)^{m_2} \\
T^{m_1m_2m_3}_{F3} &=& 2p_3^{m_1}p_1^{m_2}p_2^{m_3}-(\eta^{m_1m_2}p_1^{m_3}p_3^2
                            +\eta^{m_2m_3}p_2^{m_1}p_1^2+\eta^{m_3m_1}p_3^{m_2}p_2^2) \\
T^{m_1m_2m_3}_{F4} &=& -\Bigl(\eta^{m_2m_3}k^{m_1}[ -k^2-(k-p_1-p_2)^2]
                            +\eta^{m_1m_2}k^{m_3}[-(k-p_1)^2-(k-p_1-p_2)^2] \nonumber \\
                         & &-\eta^{m_1m_2}(k-p_1)^{m_3}[-k^2-(k-p_1-p_2)^2]
                            +\eta^{m_1m_2}k^{m_3}[-k^2-(k-p_1)^2] \nonumber \\
                         & &-\eta^{m_2m_3}(k-p_2)^{m_1}[-k^2-(k-p_1)^2]
                            +\eta^{m_3m_1}(k-p_1)^{m_2}[-k^2-(k-p_1-p_2)^2] \nonumber \\
                         & &-\eta^{m_3m_1}k^{m_2}[-(k-p_1)^2-(k-p_1-p_2)^2]
                            +\eta^{m_3m_1}(k-p_1)^{m_2}[-k^2-(k-p_1)^2] \nonumber \\
                         & &+\eta^{m_2m_3}k^{m_1}[-(k-p_1)^2-(k-p_1-p_2)^2]\Bigr).
   \label{Floop}
\end{eqnarray}

The diagram (1) provides
\begin{equation}
-32\pi^3\lambda^3E_{m_1m_2m_3}\int_k\frac1{k^2(k-p_1)^2(k-p_1-p_2)^2}[ 8k^{m_1}(k-p_1)^{m_2}k^{m_3} ],
\end{equation}
which cancels the term $T_{F1}^{m_1m_2m_3}$ in (\ref{Floop}).

The diagram (4) provides
\begin{eqnarray}
& &-4\pi^3i\lambda^3E_{m_1m_2m_3}\Bigl[\ \frac1{|p_3|}(-\eta^{m_2m_3}p_3^{m_1}
   +\eta^{m_3m_1}p_3^{m_2})+\frac1{|p_1|}(-\eta^{m_3m_1}p_1^{m_2}
   +\eta^{m_1m_2}p_1^{m_3}) \nonumber \\
& &\hspace{1cm}+\frac1{|p_2|}(-\eta^{m_1m_2}p_2^{m_3}
   +\eta^{m_2m_3}p_2^{m_1}) \Bigr]
 + (\mbox{non-covariant terms}).
   \label{NC0}
\end{eqnarray}
The covariant terms above cancel the covariant terms coming from $T_{F4}^{m_1m_2m_3}$ in (\ref{Floop}).

One can show that
\begin{equation}
-2(p_2^{m_1}p_1^{m_2}-\eta^{m_1m_2}p_1\cdot p_2)E_{m_1m_2m_3}
= 2\left[ 1-p_1^0p_2^0\frac{p_1\cdot p_2}{p_1^2p_2^2}+\frac{(p_2^0)^2}{p_2^2}+\frac{(p_1^0)^2}{p_1^2} \right]
\frac{\epsilon_{0l_3m_3}p_3^{l_3}}{p_3^2}.
\end{equation}
By using this formula, it turns out that
the term $T_{F2}^{m_1m_2m_3}$ in (\ref{Floop}) provides
\begin{eqnarray}
& & -32\pi^3\lambda^3\int_k\frac1{k^2(k-p_1)^2(k-p_1-p_2)^2}\left[
 2k^{m}\frac{\epsilon_{0l m}p_3^l}{p_3^2}+2k^m\frac{\epsilon_{0l m}p_1^l}{p_1^2}
 +2(k-p_1)^m\frac{\epsilon_{0l m}p_2^l}{p_2^2} \right] \nonumber \\
& & -32\pi^3\lambda^3\int_k\frac1{k^2(k-p_1)^2(k-p_1-p_2)^2}\left[ \left(
 -2p_1^0p_2^0\frac{p_1\cdot p_2}{p_1^2p_2^2}+2\frac{(p_2^0)^2}{p_2^2}+2\frac{(p_1^0)^2}{p_1^2} \right)
 k^{m_3}\frac{\epsilon_{0l_3m_3}p_3^{l_3}}{p_3^2} \right. \nonumber \\
& & \hspace*{5cm}  +\left( -2p_2^0p_3^0\frac{p_2\cdot p_3}{p_2^2p_3^2}+2\frac{(p_3^0)^2}{p_3^2}+2\frac{(p_2^0)^2}{p_2^2} \right)
 k^{m_1}\frac{\epsilon_{0l_1m_1}p_1^{l_1}}{p_1^2}
   \label{NC1} \\
& & \hspace*{5cm} \left. +\left( -2p_3^0p_1^0\frac{p_3\cdot p_1}{p_3^2p_1^2}+2\frac{(p_1^0)^2}{p_1^2}+2\frac{(p_3^0)^2}{p_3^2}
 \right) (k-p_1)^{m_2}\frac{\epsilon_{0l_2m_2}p_2^{l_2}}{p_2^2} \right]. \nonumber
\end{eqnarray}
The first integral is canceled by the diagrams (3).

Finally, the term $T_{F3}^{m_1m_2m_3}$ in (\ref{Floop}) provides
\begin{equation}
-32\pi^3\lambda^3\int_k\frac1{k^2(k-p_1)^2(k-p_1-p_2)^2}\left[
-\frac{\epsilon_{0l m}p_1^l p_2^m}{p_1^2p_2^2p_3^2}\left( p_1^2(p_2^0)^2+p_2^2(p_1^0)^2-2p_1\cdot p_2(p_1^0p_2^0) \right) \right].
   \label{NC2}
\end{equation}

In summary, all the covariant terms from the diagrams in figure \ref{3-3cancel} cancel among them,
and provide the terms in (\ref{NC0}), (\ref{NC1}) and (\ref{NC2}) which are non-covariant.
These non-covariant terms will vanish if the momentum and contour integrations are performed.
Namely, the contour integrals provide $\prod_{i=1}^3\delta(p_i^0)$, and therefore, those terms vanish by the momentum integration.


\begin{thebibliography}{99}

\bibitem{Maldacena:1997re}
  J.~M.~Maldacena,
  ``The large N limit of superconformal field theories and supergravity,''
  Adv.\ Theor.\ Math.\ Phys.\  {\bf 2} (1998) 231
  [Int.\ J.\ Theor.\ Phys.\  {\bf 38} (1999) 1113]
  [arXiv:hep-th/9711200].


\bibitem{Wilson:1974sk}
  K.~G.~Wilson,
  ``Confinement of quarks,''
  Phys.\ Rev.\  D {\bf 10} (1974) 2445.

\bibitem{Rey:1998ik}
  S.~J.~Rey and J.~T.~Yee,
  ``Macroscopic strings as heavy quarks in large N gauge theory and  anti-de
  Sitter supergravity,''
  Eur.\ Phys.\ J.\  C {\bf 22} (2001) 379
  [arXiv:hep-th/9803001].
  S.~J.~Rey, S.~Theisen and J.~T.~Yee,
  ``Wilson-Polyakov loop at finite temperature in large N gauge theory and
  anti-de Sitter supergravity,''
  Nucl.\ Phys.\  B {\bf 527} (1998) 171
  [arXiv:hep-th/9803135].

\bibitem{Maldacena:1998im}
  J.~M.~Maldacena,
  ``Wilson loops in large N field theories,''
  Phys.\ Rev.\ Lett.\  {\bf 80} (1998) 4859
  [arXiv:hep-th/9803002].






\bibitem{Erickson:2000af}
  J.~K.~Erickson, G.~W.~Semenoff and K.~Zarembo,
  ``Wilson loops in N = 4 supersymmetric Yang-Mills theory,''
  Nucl.\ Phys.\  B {\bf 582} (2000) 155
  [arXiv:hep-th/0003055].

\bibitem{Drukker:2000rr}
  N.~Drukker and D.~J.~Gross,
  ``An exact prediction of N = 4 SUSYM theory for string theory,''
  J.\ Math.\ Phys.\  {\bf 42} (2001) 2896
  [arXiv:hep-th/0010274].



\bibitem{Pestun:2007rz}
  V.~Pestun,
   ``Localization of gauge theory on a four-sphere and supersymmetric Wilson
  loops,''
  arXiv:0712.2824 [hep-th].


\bibitem{Aharony:2008ug}
  O.~Aharony, O.~Bergman, D.~L.~Jafferis and J.~Maldacena,
   ``N=6 superconformal Chern-Simons-matter theories, M2-branes and their
  gravity duals,''
  arXiv:0806.1218 [hep-th].


\bibitem{Witten:1988hf}
  E.~Witten,
  ``Quantum field theory and the Jones polynomial,''
  Commun.\ Math.\ Phys.\  {\bf 121} (1989) 351.

\bibitem{Guadagnini:1989am}
  E.~Guadagnini, M.~Martellini and M.~Mintchev,
  ``Wilson Lines in Chern-Simons Theory and Link Invariants,''
  Nucl.\ Phys.\  B {\bf 330} (1990) 575.

\bibitem{Berenstein:2008dc}
  D.~Berenstein and D.~Trancanelli,
  ``Three-dimensional N=6 SCFT's and their membrane dynamics,''
  arXiv:0808.2503 [hep-th].


\bibitem{Mukhi:2008ux}
  S.~Mukhi and C.~Papageorgakis,
  ``M2 to D2,''
  JHEP {\bf 0805} (2008) 085
  [arXiv:0803.3218 [hep-th]].

\bibitem{Honma:2008jd}
  Y.~Honma, S.~Iso, Y.~Sumitomo and S.~Zhang,
  ``Scaling limit of N=6 superconformal Chern-Simons theories and Lorentzian
  Bagger-Lambert theories,''
  arXiv:0806.3498 [hep-th].

\bibitem{Li:2008ya}
  T.~Li, Y.~Liu and D.~Xie,
  ``Multiple D2-Brane Action from M2-Branes,''
  arXiv:0807.1183 [hep-th].

\bibitem{Pang:2008hw}
  Y.~Pang and T.~Wang,
  ``From N M2's to N D2's,''
  arXiv:0807.1444 [hep-th].

\bibitem{Honma:2008ef}
  Y.~Honma, S.~Iso, Y.~Sumitomo, H.~Umetsu and S.~Zhang,
  ``Generalized Conformal Symmetry and Recovery of SO(8) in Multiple M2 and D2
  Branes,''
  arXiv:0807.3825 [hep-th].



\bibitem{Bagger:2007jr}
  J.~Bagger and N.~Lambert,
  ``Gauge Symmetry and Supersymmetry of Multiple M2-Branes,''
  Phys.\ Rev.\  D {\bf 77} (2008) 065008
  [arXiv:0711.0955 [hep-th]].


\bibitem{Gustavsson:2007vu}
  A.~Gustavsson,
  ``Algebraic structures on parallel M2-branes,''
  arXiv:0709.1260 [hep-th].

\bibitem{Bak:2008cp}
  D.~Bak and S.~J.~Rey,
  ``Integrable Spin Chain in Superconformal Chern-Simons Theory,''
  arXiv:0807.2063 [hep-th].

\bibitem{Drukker:2008zx}
  N.~Drukker, J.~Plefka and D.~Young,
  ``Wilson loops in 3-dimensional N=6 supersymmetric Chern-Simons Theory and
  their string theory duals,''
  arXiv:0809.2787 [hep-th].

\bibitem{Chen:2008bp}
  B.~Chen and J.~B.~Wu,
  ``Supersymmetric Wilson Loops in N=6 Super Chern-Simons-matter theory,''
  arXiv:0809.2863 [hep-th].

\bibitem{Kluson:2008wn}
  J.~Kluson and K.~L.~Panigrahi,
  ``Defects and Wilson Loops in 3d QFT from D-branes in AdS(4) x CP**3,''
  arXiv:0809.3355 [hep-th].


\bibitem{Berenstein:1998ij}
  D.~E.~Berenstein, R.~Corrado, W.~Fischler and J.~M.~Maldacena,
  ``The operator product expansion for Wilson loops and surfaces in the  large
  N limit,''
  Phys.\ Rev.\  D {\bf 59} (1999) 105023
  [arXiv:hep-th/9809188].

\bibitem{Drukker:1999zq}
  N.~Drukker, D.~J.~Gross and H.~Ooguri,
  ``Wilson loops and minimal surfaces,''
  Phys.\ Rev.\  D {\bf 60} (1999) 125006
  [arXiv:hep-th/9904191].


\bibitem{Zarembo:2002an}
  K.~Zarembo,
  ``Supersymmetric Wilson loops,''
  Nucl.\ Phys.\  B {\bf 643} (2002) 157
  [arXiv:hep-th/0205160].

\bibitem{Dymarsky:2006ve}
  A.~Dymarsky, S.~S.~Gubser, Z.~Guralnik and J.~M.~Maldacena,
  ``Calibrated surfaces and supersymmetric Wilson loops,''
  JHEP {\bf 0609} (2006) 057
  [arXiv:hep-th/0604058].


\bibitem{mine}
S.-J. Rey, unpublished notes (2004, 2008); to appear.

\bibitem{miwa}
A.~Miwa,
  ``BMN operators from Wilson loop,''
  JHEP {\bf 0506} (2005) 050
  [arXiv:hep-th/0504039].

\bibitem{fujita}
M.~Fujita,
  ``BPS operators from the Wilson loop in the 3-dimensional supersymmetric
  Chern-Simons theory,''
  arXiv:0902.2381 [hep-th].


\bibitem{Gaiotto:2008cg}
  D.~Gaiotto, S.~Giombi and X.~Yin,
  ``Spin Chains in N=6 Superconformal Chern-Simons-Matter Theory,''
  arXiv:0806.4589 [hep-th].

\bibitem{Hosomichi:2008jb}
  K.~Hosomichi, K.~M.~Lee, S.~Lee, S.~Lee and J.~Park,
  ``N=5,6 Superconformal Chern-Simons Theories and M2-branes on Orbifolds,''
  arXiv:0806.4977 [hep-th].

\bibitem{Terashima:2008sy}
  S.~Terashima,
  ``On M5-branes in N=6 Membrane Action,''
  JHEP {\bf 0808} (2008) 080
  [arXiv:0807.0197 [hep-th]].

\bibitem{Gaiotto:2007qi}
  D.~Gaiotto and X.~Yin,
  ``Notes on superconformal Chern-Simons-matter theories,''
  JHEP {\bf 0708}, 056 (2007)
  [arXiv:0704.3740 [hep-th]].

\bibitem{Chen:1992ee}
  W.~Chen, G.~W.~Semenoff and Y.~S.~Wu,
  ``Two loop analysis of nonAbelian Chern-Simons theory,''
  Phys.\ Rev.\  D {\bf 46}, 5521 (1992)
  [arXiv:hep-th/9209005].

\bibitem{Nishioka:2008gz}
  T.~Nishioka and T.~Takayanagi,
  ``On Type IIA Penrose Limit and N=6 Chern-Simons Theories,''
  JHEP {\bf 0808}, 001 (2008)
  [arXiv:0806.3391 [hep-th]].

\bibitem{Grignani:2008is}
  G.~Grignani, T.~Harmark and M.~Orselli,
  ``The SU(2) x SU(2) sector in the string dual of N=6 superconformal
  Chern-Simons theory,''
  arXiv:0806.4959 [hep-th].

\end{thebibliography}
\end{document}